\documentclass[aps,prd,10pt,notitlepage,superscriptaddress,nofootinbib,numbers]{revtex4-2}

\usepackage[utf8]{inputenc}
\usepackage[T1]{fontenc}
\usepackage{lmodern}
\usepackage{microtype}
\usepackage{amsmath,amssymb,amsfonts,mathrsfs}
\usepackage{bm}
\usepackage{graphicx}
\usepackage{float}
\usepackage[dvipsnames]{xcolor}
\usepackage{hyperref}
\hypersetup{colorlinks=true,citecolor=Purple,linkcolor=Purple,urlcolor=Purple,breaklinks=true}
\usepackage[capitalize]{cleveref}
\usepackage{tikz}

\definecolor{lime}{HTML}{A6CE39}
\DeclareRobustCommand{\orcidicon}{%
	\begin{tikzpicture}
		\draw[lime, fill=lime] (0,0) 
		circle [radius=0.16] 
		node[white] {{\fontfamily{qag}\selectfont \tiny ID}};
		\draw[white, fill=white] (-0.0625,0.095) 
		circle [radius=0.007];
	\end{tikzpicture}
	\hspace{-2mm}
}

\foreach \x in {A, ..., Z}{%
	\expandafter\xdef\csname orcid\x\endcsname{\noexpand\href{https://orcid.org/\csname orcidauthor\x\endcsname}{\noexpand\orcidicon}}
}

\newcommand\orcidJonathan{{\href{https://orcid.org/0000-0001-9291-0893}{\orcidicon}}}
\newcommand\orcidEdson{{\href{https://orcid.org/0000-0001-9929-5977}{\orcidicon}}}

\begin{document}

\title{Optical appearance of radially stable quantum-improved thin-shell wormholes supported by cosmic Chaplygin gases}
\author{Jonathan A. Rebou\c{c}as\orcidJonathan\!\!}
\email{jalvesreboucas@ifce.edu.br}
\affiliation{Instituto Federal de Educa\c{c}\~ao, Ci\^encia e Tecnologia do Cear\'a (IFCE), Iguatu, CE, Brazil}

\author{Edson Otoniel\orcidEdson\!\!}
\email{edson.otoniel@ufca.edu.br}
\affiliation{Universidade Federal do Cariri (UFCA), Instituto de Forma\c{c}\~ao de Educadores - IFE,  R. Oleg\'ario Emidio de Araujo S/N, Brejo Santo - CE, 63.260-000 - Brazil}
\begin{abstract}
We study the optical appearance and radial support of symmetric thin-shell wormholes built from quantum-improved Schwarzschild exteriors. Two proper-time prescriptions of asymptotically safe gravity, B and C, supply the running Newton couplings used to improve the metric with a common infrared scale identification. Their distinct treatments of gravitational fluctuations enter classical light propagation through the effective metric. The supporting generalized and modified cosmic Chaplygin gases are phenomenological surface fluids, with real negative-density branches and material calibration fixed during each perturbation. In the respective parameter slices studied, the generalized gas admits throats with linear radial stability between the seed photon sphere and innermost stable circular orbit, whereas stable modified-gas sectors lie below the photon sphere. At equal mass and normalized deformation, C shifts the orbital and critical scales farther inward than B. Ray tracing of celestial sources and orbiting disks resolves this prescription dependence. For a matched outer cut, C loses a celestial source-image order accessible in B. With an inclined observer-side disk truncated at the innermost stable circular orbit, a throat below the photon sphere reproduces the image of the black hole with the same exterior. An outer cut can remove returning disk contributions, while illuminating a second disk adds internal arcs along transmitted rays. A central dark region can persist without a horizon. Radially supported examples realize both optical regimes. The combined analysis connects the quantum-running dependence of the geometry to material viability and source-dependent image structure.
\end{abstract}

\maketitle
\tableofcontents
\raggedbottom

\section{Introduction}\label{sec:introduction}
A quantum description of gravitation must connect the successful macroscopic dynamics of spacetime to the behavior of quantum fields and geometry at short distances. Black-hole entropy and particle creation already link causal boundaries, thermodynamics and quantum theory, without by themselves specifying a complete microscopic gravitational theory~\cite{Bekenstein1973,Hawking1975}. At low energies, general relativity admits a controlled effective-field-theory treatment that separates calculable quantum contributions from information about its high-energy completion~\cite{Donoghue1994,Burgess2004}. Extending that description into the ultraviolet (UV), and identifying which consequences could remain accessible in macroscopic strong-field systems, are distinct parts of the quantum-gravity problem. Effective spacetime models are useful in this connection when their quantum input, additional assumptions and observable predictions remain distinguishable.

Asymptotically safe gravity (ASG) explores a quantum-field-theoretic completion in which the renormalization group (RG) approaches an interacting fixed point with a finite number of relevant directions. The effective average action provides a nonperturbative framework for investigating this possibility, with scale-dependent gravitational couplings obtained from a functional evolution equation~\cite{Reuter1998,Niedermaier2007,ReuterSaueressig2012}. Fixed-point evidence has been developed in Einstein--Hilbert and higher-curvature truncations, using different implementations of the coarse graining~\cite{Litim2004,Codello2009,Benedetti2009}. These results motivate a controlled investigation of quantum-running effects, while their truncation and regulator dependence must be retained when connecting the flow to a particular effective geometry.

The interpretation of this evidence also requires more than locating a zero of a set of beta functions. Gravity--matter systems, the distinction between background and fluctuating metrics, and the relation between UV behavior and a classical infrared (IR) regime are central ingredients of the program~\cite{Eichhorn2019,PawlowskiReichert2021}. Critical assessments identify open questions concerning background independence, unitarity, signature and the construction of observables, while recent reviews describe the expanding phenomenological applications~\cite{Bonanno2020Critical,Eichhorn2026Review}. A comparison between different coarse-graining prescriptions can therefore test the dependence of an effective prediction on its quantum input. It does not automatically isolate a regulator-independent observable of a completed theory.

Black holes offer a concrete setting for making that connection. RG improvement of a classical solution replaces selected couplings by their running counterparts, whereas improvement at the action level changes the equations obtained through variation~\cite{BonannoReuter2000,ReuterWeyer2004}. A spacetime-dependent coupling additionally requires a scale identification relating the RG description to geometric quantities. Reviews of quantum-improved black holes show how this choice enters causal structure, collapse, thermodynamics and other effective properties~\cite{KochSaueressig2014,Platania2023,Spina2025Review}. Optical applications further illustrate that an assumed scale for new gravitational effects can influence a predicted shadow~\cite{Held2019}. Thus a running coupling supplies quantum information to a model, but its insertion into a metric is a further physical prescription whose domain must be stated.

Proper-time flows provide a useful setting for examining these choices. Regulator, gauge and metric-parametrization dependence have been investigated directly in quantum Einstein gravity, and related running laws have been applied to black-hole models and their perturbative and optical properties~\cite{BonannoOglialoroZappala2025,BonannoEtAl2025}. The present work compares the two prescriptions denoted B and C in that construction. For their spacetime implementation, we adopt the IR distance prescription used in a quantum-improved Schwarzschild study of thermodynamics and shadows~\cite{Muniz2026}. A complementary wormhole application inserts a running coupling into field equations with a Dekel--Zhao matter distribution~\cite{Reboucas2026DZ}. The latter has a different bulk source and closure. Our construction instead uses prescribed improved exteriors and a thin material junction, so neither that halo matter nor its field-equation interpretation is transferred to the shell.

Traversable wormholes test a different aspect of an effective gravitational description: two exterior regions can communicate through a throat without possessing the causal structure of a black hole. Explicit traversable geometries and surgical constructions make clear that such a connection requires a simultaneous account of geometry and supporting stresses~\cite{MorrisThorne1988,Visser1989Surgery}. A modified exterior alone does not specify the matter needed to maintain the throat or its response to a disturbance. This distinction is particularly relevant to quantum-improved geometries, for which a quantum origin of the running couplings need not imply that the junction matter has been derived from a quantum state.

The energy requirements constrain this interpretation. Local throat analyses relate traversability to violations of the null energy condition (NEC), while topological-censorship results impose restrictions under their stated asymptotic, causal and energy assumptions~\cite{HochbergVisser1998,Friedman1993}. Measures of integrated exotic matter can become small in certain constructions, but that observation does not remove the conditions on the local stress tensor~\cite{VisserKarDadhich2003}. If negative energy is attributed to quantum fields, quantum inequalities provide additional restrictions on its magnitude, duration and geometric scales~\cite{FordRoman1996,FewsterRoman2005}. These field-theoretic restrictions are separate from assuming a nonlinear surface-fluid law. The distributional energy conditions of a phenomenological shell are therefore distinct from a microscopic explanation of its supporting material.

The thin-shell framework makes that distinction explicit by joining two exterior domains across a timelike hypersurface. The jump in extrinsic curvature determines the required surface stresses through the junction formalism~\cite{Israel1966}. Radial equilibrium and restoring behavior then depend on both the exterior geometry and the shell equation of state (EOS), as demonstrated by linearized stability analyses and their general spherical extensions~\cite{PoissonVisser1995,LoboCrawford2004,GarciaLoboVisser2012}. A static junction is not automatically stable. Moreover, calibrating a material parameter to obtain an equilibrium does not permit that parameter to change during the perturbation used to test it. Comparing backgrounds at matched geometric and free material inputs must retain this distinction between a family of separately calibrated equilibria and the dynamics of each member.

Chaplygin-type fluids offer nonlinear constitutive models for this comparison. Their cosmological development includes the original and generalized Chaplygin descriptions of accelerated expansion and matter--dark-energy interpolation~\cite{Kamenshchik2001,Bento2002}. The generalized cosmic Chaplygin gas (GCCG) and modified cosmic Chaplygin gas (MCCG) extend that family, with the latter also admitting a viscous cosmological completion~\cite{GonzalezDiaz2003,Pourhassan2013}. Here their role is a surface EOS rather than a homogeneous cosmological fluid: only the stated nondissipative constitutive relations are adopted. Their use at negative shell density requires a specified real branch and a fixed dimensional normalization. Cosmological motivation does not determine those surface choices or establish a microscopic quantum theory of the material.

Applications of Chaplygin laws to spherical thin shells have shown how charge, a cosmological constant and nonlinear electromagnetic backgrounds enter the radial stability problem~\cite{EiroaSimeone2007,Eiroa2009,EiroaFigueroa2012}. Modified and cosmic variants have also been investigated as supporting matter in related constructions~\cite{SharifAzam2013,Azam2016Cosmic}. These studies motivate comparing constitutive responses, while their individual stable sectors remain conditional on the geometry, density branch and material parameters used. They do not supply a stability ordering that can be transferred unchanged to B and C. In a quantum-running comparison, the changed exterior domain and the changed pressure response must both be included before attributing a restoring sector to the gravitational improvement.

The optical problem supplies an independent test of these candidate geometries. Horizon-scale imaging of M87 has made source-weighted strong-field appearance an observationally relevant question, alongside the broader search for dark compact-object alternatives~\cite{EHT2019M87,CardosoPani2019}. However, a bright ring or a central intensity depression does not identify a causal boundary without an emission model. Analytic studies distinguish photon-critical structure from broader lensing features and show how emissivity and frequency transport shape an image~\cite{GrallaHolzWald2019}. Photon subrings can carry geometric information in interferometric observables, but that information is different from the full resolved brightness distribution~\cite{Johnson2020,GrallaLupsascaMarrone2020}. A comparison with Schwarzschild must consequently control the observer, emitting support and source motion, rather than identify every luminous annulus with a photon sphere.

Wormhole imaging makes the role of the source particularly clear. Calculations for traversable geometries with emitting matter demonstrate that ring contrast and internal illumination depend on the adopted radiation environment~\cite{Bambi2013,OhgamiSakai2015}. In asymmetric thin-shell and two-region geometries, transmitted or returned rays can generate additional image branches controlled by the throat and the null structures on either side~\cite{Wang2020,PengGuoFeng2021,Huang2023}. The compatibility of such configurations with radial support is a separate issue, also examined for reflection-asymmetric examples~\cite{Tsukamoto2021}. These results provide useful mechanisms for comparison, but geometric asymmetry and different source prescriptions prevent their individual signatures from being assigned directly to a symmetric quantum-improved shell.

Recent thin-shell studies with regularized string-cloud and exponential-electrodynamic seeds combine material stability with celestial and moving-disk diagnostics~\cite{Zhong2026StringShell,Reboucas2026ExponentialTSW}. Related asymmetric calculations separate changes in the retained exterior from the effects of source support and emitter kinematics~\cite{Zhong2026AsymmetricOptics}. Their common lesson is that regularity of a removed core, existence of a static throat and appearance of an illuminated image answer different questions. In particular, an observer-side disk need not distinguish a symmetric wormhole from the black hole with the same exterior, whereas emission in the opposite region can activate additional intersections. This source dependence motivates testing the two ASG prescriptions with matched illumination before interpreting their image differences as consequences of quantum gravitational running.

We construct symmetric thin-shell wormholes from the B and C improved Schwarzschild exteriors and compare them with the Schwarzschild limit at matched mass and normalized deformation. The junction stresses and local radial response are evaluated with GCCG and MCCG on their declared real branches, calibrating equilibrium once and holding the resulting material inputs fixed during each perturbation. We then calculate null propagation, celestial maps and inclined disk images for cuts on either side of the seed photon sphere. The emitting disks are truncated at the innermost stable circular orbit (ISCO), and unilateral and bilateral source choices are compared separately. The aim is to connect the quantum-running dependence of the retained geometry to the available restoring sectors and source-weighted optical distinctions. The exteriors remain effective IR-inspired prescriptions, the shell material is phenomenological, and light follows classical geometric optics; the analysis does not identify a microscopic quantum shell or a unique observable of ASG.

This paper is organized as follows. Section~\ref{sec:asg} presents asymptotic safety and the improvement prescriptions. The seed geometries and their effective interpretation are described in Sec.~\ref{sec:geometry}. Sections~\ref{sec:junction} and~\ref{sec:stability} develop the symmetric junction and radial stability analysis, respectively. Section~\ref{sec:eos} applies the GCCG and MCCG laws, and Sec.~\ref{sec:optics} presents the emission model, imaging method and optical results. Section~\ref{sec:conclusion} summarizes the conclusions.

\section{Asymptotic safety and improvement prescriptions}\label{sec:asg}

\subsection{Scale dependence and the gravitational fixed point}\label{subsec:asg-flow}

Reuter's functional RG formulation of ASG describes a possible UV completion through a non-Gaussian fixed point with finitely many relevant directions~\cite{Reuter1998}. In natural units \(c=\hbar=1\), where \(c\) is the speed of light and \(\hbar\) the reduced Planck constant, the Newton coupling has dimensions of length squared. Its dimensionless form satisfies
\begin{equation}
 \mathfrak g_X(\varkappa)=\varkappa^2G_X(\varkappa),
 \qquad
 \lim_{\varkappa\to\infty}\mathfrak g_X(\varkappa)=g_{*,X}.
 \label{eq:asg-fixed-point}
\end{equation}
Here \(\varkappa\geq0\) is the RG momentum scale, of inverse-length dimension, \(G_X\) is the running Newton coupling for prescription \(X\), \(\mathfrak g_X\) its dimensionless counterpart, and \(g_{*,X}>0\) its fixed-point coordinate. The normalization is \(G_X(0)=G_0>0\), with \(G_0\) the IR Newton constant. The running couplings are coefficients of a coarse-grained gravitational effective action: their scale dependence incorporates quantum fluctuations as the RG scale is lowered~\cite{Reuter1998}. The fixed point concerns this quantum flow, while its spacetime consequences additionally depend on how the RG scale is assigned to the geometry.

Fixed-point coordinates depend on the truncation, regulator, gauge and metric parametrization~\cite{BonannoOglialoroZappala2025}. A spacetime application additionally specifies a physical scale identification; the UV limit alone does not determine that prescription.

\subsection{Action, field-equation and solution improvement}\label{subsec:asg-implementations}

Reuter and Weyer~\cite{ReuterWeyer2004} introduce position-dependent couplings in the gravitational action before metric variation. Their modified Einstein--Hilbert term, restricted here to a vanishing cosmological coupling \(\Lambda_{\mathrm c}=0\), is
\begin{equation}
 S_{\mathrm{grav}}=\frac{1}{16\pi}\int d^4z\,\sqrt{-g}\,
 \frac{\mathcal R}{G(\varkappa(z))}.
 \label{eq:asg-action}
\end{equation}
The dimensionless \(S_{\mathrm{grav}}\) is the gravitational action, \(z\) denotes spacetime coordinates \(z^\mu\), \(g\) is the metric determinant, and \(\mathcal R\) is the Ricci scalar. Greek indices label four spacetime coordinates; \(\Lambda_{\mathrm c}\) and \(\mathcal R\) have inverse-length-squared dimensions. The assigned \(\varkappa(z)\) determines the position-dependent coupling \(G\). With that assignment held fixed, metric variation produces derivatives of \(G^{-1}\); further terms depend on the specified coupling dynamics and boundary action~\cite{ReuterWeyer2004}.

At the field-equation level, Rebou\c{c}as et al.~\cite{Reboucas2026DZ} insert a radial Newton coupling into the bulk equations of a Dekel--Zhao halo wormhole. With \(\Lambda_{\mathrm c}=0\), their prescription can be written in geometrized source normalization as
\begin{equation}
 \mathcal G_{\mu\nu}
 =8\pi\,\frac{G}{G_0}\,\mathcal T_{\mu\nu}.
 \label{eq:asg-field-improvement}
\end{equation}
Here \(\mathcal G_{\mu\nu}\) is the Einstein tensor and \(\mathcal T_{\mu\nu}\) is the physical material stress tensor multiplied by the fixed \(G_0\); both have inverse-length-squared dimensions in an orthonormal frame. The Bianchi identity requires \(\nabla^\mu[(G/G_0)\mathcal T_{\mu\nu}]=0\), where \(\nabla_\mu\) is the metric-compatible covariant derivative. Separate material conservation therefore requires an additional consistency check.

Bonanno and Reuter~\cite{BonannoReuter2000} instead improve the Schwarzschild solution through the substitution \(G_0/r\mapsto G(\varkappa(r))/r\) in its potential, where \(r\) is the areal radius. Their effective-matter interpretation expresses the resulting geometry through Einstein equations with constant \(G_0\). We adopt this solution improvement for the exterior metrics and that effective description for the shell, using the Einstein--Israel junction law~\cite{BonannoReuter2000,Israel1966}. The running changes the geometry; the junction coupling remains \(G_0\). This is a specified effective closure, whose junction law is not derived from \cref{eq:asg-action}.

\subsection{Two proper-time running prescriptions}\label{subsec:asg-schemes}

Bonanno, Oglialoro and Zappal\`a~\cite{BonannoOglialoroZappala2025} distinguish proper-time regulator prescriptions B and C. Scheme B retains the scale derivative of the running kinetic normalization; C uses an alternative regulator. They implement distinct coarse-graining prescriptions for quantum gravitational fluctuations within a proper-time approximation. Comparing them tests how that choice propagates into effective spacetime predictions~\cite{BonannoOglialoroZappala2025}. For the reduced four-dimensional Newton flow at \(\Lambda_{\mathrm c}=0\), Bonanno et al.~\cite{BonannoEtAl2025} obtain
\begin{equation}
 \frac{G_{\mathrm B}(\varkappa)}{G_0}=\left[\frac{\varkappa^2G_0}{2g_{*,\mathrm B}}+\sqrt{1+\frac{\varkappa^4G_0^2}{4g_{*,\mathrm B}^{\,2}}}\right]^{-1},\qquad
 \frac{G_{\mathrm C}(\varkappa)}{G_0}=\left(1+\frac{\varkappa^2G_0}{g_{*,\mathrm C}}\right)^{-1}.
 \label{eq:asg-momentum-couplings}
\end{equation}
The labels B and C specify \(X\), with their respective \(g_{*,X}\). Both positive functions approach \(G_0\) at zero momentum and \(g_{*,X}/\varkappa^2\) in the UV. Their fixed-point normalization can be absorbed into the cutoff amplitude, while their interpolation remains different~\cite{BonannoEtAl2025}. The zero reference momentum is distinct from the assumption \(\Lambda_{\mathrm c}=0\).

The regular black holes of Ref.~\cite{BonannoEtAl2025} also use a matter-dependent cutoff and collapse matching; regularity does not follow from the momentum functions alone. Our exterior model adopts the IR distance prescription of Muniz et al.~\cite{Muniz2026}.

\section{Seed geometries and effective gravitational description}\label{sec:geometry}

For either running prescription, the static, spherically symmetric improved Schwarzschild exterior is written as
\begin{equation}
 ds_X^2=-f_X(r)\,dt^2+\frac{dr^2}{f_X(r)}
       +r^2\left(d\theta^2+\sin^2\theta\,d\phi^2\right).
 \label{eq:metric}
\end{equation}
The interval \(ds_X^2\) has signature \((-+++)\). The coordinate \(t\) is static time, the areal radius satisfies \(r>0\), and \(\theta\in[0,\pi]\) and \(\phi\in[0,2\pi)\) are the polar and azimuthal angles. The lapse \(f_X(r)\) is dimensionless. Time and radius have dimensions of length in the units already specified.

To implement solution improvement, let \(\varkappa_X(r)\) be the RG scale assigned to radius \(r\) for the selected prescription. Replacing the Newton factor in the classical Schwarzschild potential gives
\begin{equation}
 f_X(r)=1-\frac{2M}{r}\,
             \frac{G_X(\varkappa_X(r))}{G_0}.
 \label{eq:metric-improvement}
\end{equation}
The constant \(M>0\) is the geometric mass, with dimensions of length; it equals the physical seed mass multiplied by \(G_0\) in the natural units used for the flow. The ratio in \cref{eq:metric-improvement} is the precise place at which the running changes the metric. This solution-level substitution follows the Schwarzschild improvement strategy of Ref.~\cite{BonannoReuter2000}; it is made before constructing the shell.

We adopt the normalized cutoff identification of Ref.~\cite{Muniz2026},
\begin{equation}
 \varkappa_X(r)\sqrt{\frac{G_0}{g_{*,X}}}
 =\frac{\ell}{\mathcal D(r)}.
 \label{eq:asg-radial-scale}
\end{equation}
Here \(\mathcal D(r)>0\) is the proper-distance-based length scale entering the cutoff prescription, and \(\ell\geq0\) is a constant deformation length. The symbol \(\ell\) replaces the cutoff-distance parameter \(\xi\) of Ref.~\cite{Muniz2026}; \(\mathcal D\) replaces its distance function \(d\). The fixed-point normalization and the free cutoff amplitude are absorbed into \(\ell\). At \(\ell=0\), the prescription is understood through its continuous classical limit. Through \cref{eq:asg-radial-scale}, \(\ell\) combines the cutoff amplitude, the Newton scale, and the fixed-point normalization. Its numerical ratio to \(M\) specifies an effective deformation; assigning it to a particular physical mass requires a microscopic scale identification beyond the parameter scan.

In the IR distance approximation \(\mathcal D(r)\simeq r\), \cref{eq:asg-radial-scale} gives \(\varkappa_X^2G_0/g_{*,X}\simeq\ell^2/r^2\). Substitution into \cref{eq:asg-momentum-couplings}, with \(G_X(r)\) denoting the composition \(G_X(\varkappa_X(r))\), yields
\begin{equation}
 \frac{G_{\mathrm B}(r)}{G_0}
 =\left[\frac{\ell^2}{2r^2}
        +\sqrt{1+\frac{\ell^4}{4r^4}}\right]^{-1},
 \qquad
 \frac{G_{\mathrm C}(r)}{G_0}
 =\left(1+\frac{\ell^2}{r^2}\right)^{-1}.
 \label{eq:couplings}
\end{equation}
These are dimensionless radial ratios. Taking \(\mathcal D(r)=r\) as the effective exterior prescription makes their displayed functional forms exact within that model. The scheme-C ratio has the same radial dependence as the coupling used in the field-equation construction of Ref.~\cite{Reboucas2026DZ}. The present geometry is obtained by inserting it into \cref{eq:metric-improvement}; the Dekel--Zhao matter profile and its bulk field equations are not imposed.

Applying the two radial ratios to that lapse rule gives the explicit pair
\begin{equation}
 f_{\mathrm B}(r)=1-\frac{4Mr}{\ell^2+\sqrt{4r^4+\ell^4}},
 \qquad
 f_{\mathrm C}(r)=1-\frac{2Mr}{r^2+\ell^2}.
 \label{eq:lapses}
\end{equation}
The scheme labels retain the proper-time meaning specified in \cref{subsec:asg-schemes}. Equal values of \(M\) and \(\ell\) compare the geometries at a common mass and normalized deformation length. Matching unrescaled microscopic momentum assignments would additionally require specifying the scheme-dependent fixed-point normalization. For the spacetime and surface stresses we use geometrized normalization, with the fixed \(G_0\) absorbed, equivalently \(G_0=1\).

Both functions approach the Schwarzschild lapse continuously:
\begin{equation}
 \lim_{\ell\to0}f_X(r)=f_{\mathrm S}(r)
 =1-\frac{2M}{r}.
 \label{eq:schwarzschild}
\end{equation}
The label \(\mathrm S\) denotes the Schwarzschild control geometry. At large radius, their leading deviations are
\begin{equation}
 f_{\mathrm B}(r)=1-\frac{2M}{r}+\frac{M\ell^2}{r^3} +O\!\left(\frac{M\ell^4}{r^5}\right),\qquad f_{\mathrm C}(r)=1-\frac{2M}{r}+\frac{2M\ell^2}{r^3} +O\!\left(\frac{M\ell^4}{r^5}\right).
 \label{eq:asymptotic}
\end{equation}
The notation \(O(\cdot)\) denotes the asymptotic order of the omitted terms as \(r/\ell\) becomes large. The unequal coefficients distinguish the two geometries at fixed deformation length. The scale identification used in these metrics is an infrared approximation. Its use at arbitrarily short distances would require additional input and is outside the effective interpretation adopted here.

The seed horizons delimit the exterior retained in the construction. For prescription \(\mathrm C\), their radii are
\begin{equation}
 r_{\mathrm C,\pm}=M\pm\sqrt{M^2-\ell^2},
 \qquad 0<\ell\leq M .
 \label{eq:horizon-C}
\end{equation}
The signs \(+\) and \(-\) in this equation distinguish the outer and inner seed horizons. At \(\ell=0\), the positive Schwarzschild horizon is \(2M\); the zero-radius root of the polynomial equation is not an additional horizon. For prescription \(\mathrm B\), each positive seed horizon satisfies
\begin{equation}
 r^3-4M^2r+2M\ell^2=0,
 \qquad 4Mr\geq\ell^2 .
 \label{eq:horizon-B}
\end{equation}
The inequality retains the unsquared lapse equation and excludes spurious roots. The degenerate positive root is \(2M/\sqrt{3}\), with \(\ell^2=8M^2/(3\sqrt{3})\). Prescription \(\mathrm C\) is degenerate at \(\ell=M\). A common nonextremal black-hole-seed domain is \(0\leq\ell/M<1\), where both models possess an outer positive horizon, denoted \(r_{X,+}\). The throat is strictly outside it. Horizonless seeds and degenerate endpoints are outside this shared comparison domain. This mathematical exterior restriction does not itself guarantee infrared validity; an ASG interpretation also requires a radial regime in which the adopted distance approximation is justified.

\begin{figure}[tbp]
 \centering
 \includegraphics[width=\linewidth]{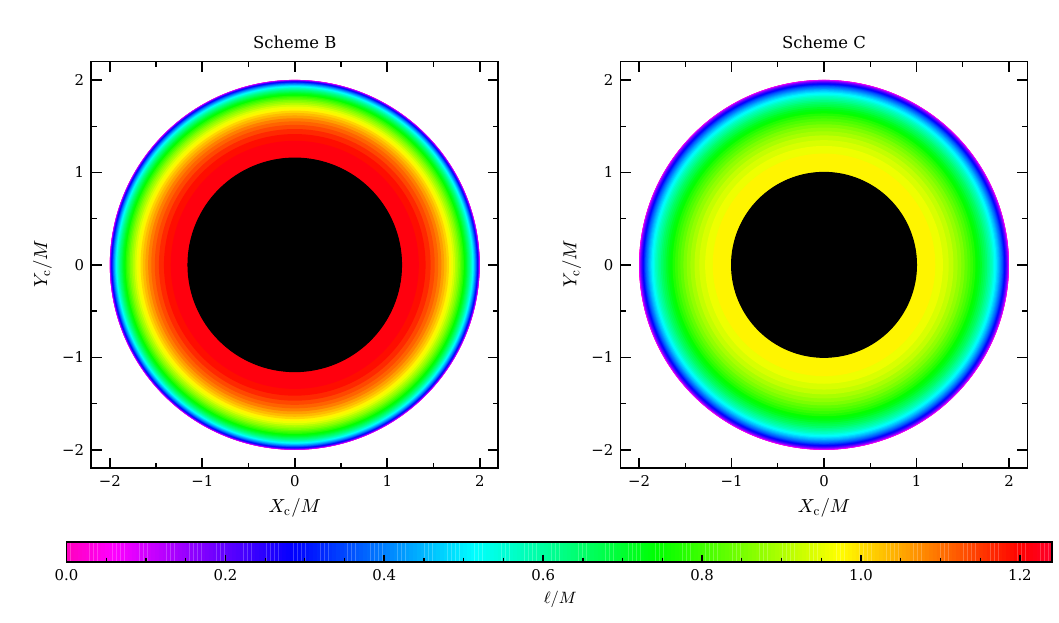}
 \caption{Outer-horizon parameter families of the prescribed ASG-inspired Schwarzschild seed metrics: scheme B for $0\leq\ell/M\leq\sqrt{8/(3\sqrt{3})}$ and scheme C for $0\leq\ell/M\leq1$. Each panel samples 49 equally spaced values including zero and the extremal endpoint, with $M=1$. The shared horizontal colorbar gives $\ell/M$ on one common scale for both panels; scheme C ends at $\ell/M=1$. The auxiliary equatorial coordinates are $X_{\rm c}=r\cos\phi$ and $Y_{\rm c}=r\sin\phi$, distinct from the prescription label $X$. At zero, the single positive horizon is $2M$; each interior parameter value has two distinct positive horizons, of which only the outer is drawn; at the upper endpoint they merge. The black core marks the minimum outer radius in the displayed family and is a graphical convention. The colored disks represent different parameter values. These are seed horizons; a thin-shell throat lies outside the retained seed horizon. This full seed scan is distinct from the common nonextremal shell-comparison domain.}
 \label{fig:asg-extremal-outer-horizons}
\end{figure}

The inward displacement of the outer horizon in Fig.~\ref{fig:asg-extremal-outer-horizons} is monotonic with \(\ell/M\), as follows by differentiating \cref{eq:horizon-C,eq:horizon-B} along their outer branches at fixed \(M\). In the common interval \(0<\ell/M<1\), scheme C has the smaller outer horizon. This ordering follows from \cref{eq:couplings}: at equal positive \(r\) and \(\ell\), the B coupling exceeds the C coupling. The lapse prescription in \cref{eq:metric-improvement} therefore places the B lapse below the C lapse. At the C outer horizon, the B lapse is still negative and must cross zero farther out before approaching its asymptotic value. Consequently, at equal mass and deformation length, the geometric lower bound on an exterior throat radius is larger for B. This displacement traces the suppression of the effective Newton coupling by its quantum running: at the same normalized cutoff, C gives the stronger reduction from \(G_0\).

The different terminal contours reflect the distinct degeneracy conditions of \cref{eq:horizon-C,eq:horizon-B}. Scheme B retains a seed horizon beyond the extremal deformation of C, so the smallest circles in the two panels correspond to different values of \(\ell/M\). Their black cores compare each prescription at its own extremal limit; equal colors on the shared scale instead identify equal deformation lengths at fixed mass. The continuation of the B family past \(\ell/M=1\) describes its additional seed-horizon domain. The degenerate endpoints shown as geometric limits remain excluded from the common nonextremal shell comparison specified above.

The nonextremal two-horizon structure also occurs in the magnetic seed used by Rebou\c{c}as et al.~\cite{Reboucas2026ExponentialTSW}, whose inner and outer horizons merge at a critical deformation. Its zero-bare-mass branch additionally has a locally Minkowski core. This regularity is a separate property from the existence of the horizon pair. For the present lapses in \cref{eq:lapses}, formal continuation to \(r\to0\) at fixed \(\ell>0\) instead gives \(\mathcal R\sim12M/(\ell^2r)\), so their approach to unit lapse does not regularize the center. This central limit lies outside the infrared justification of the adopted scale identification and supplies no ultraviolet completion. Placing the throat outside the outer seed horizon removes both horizons and the central region, while retaining the distributional matter layer at the junction.

The constant-\(G_0\) effective description specified in \cref{subsec:asg-implementations} interprets the Einstein tensor of this prescribed exterior as an effective bulk source, following the effective-matter construction of Ref.~\cite{BonannoReuter2000}. We adopt the Einstein--Israel junction law for the distributional surface layer~\cite{Israel1966}. The ASG dependence is encoded in \(f_X\) and in the resulting junction curvatures; the surface coupling is the fixed infrared constant. An action-improved theory with additional derivative or boundary terms would require its own junction derivation. The bulk parameters \(M\) and \(\ell\) remain fixed under every shell perturbation.

The same quantum-running profiles determine the characteristic radii relevant to null rays and circular disk emitters. For either seed, with a prime denoting a radial derivative, the outer unstable circular null orbit satisfies
\begin{equation}
 r_{\mathrm{ph},X}f'_X(r_{\mathrm{ph},X})-2f_X(r_{\mathrm{ph},X})=0.
 \label{eq:seed-photon-orbit}
\end{equation}
Here \(r_{\mathrm{ph},X}\) is the seed photon-sphere radius in its outer static region. It is a local maximum of the null potential \(f_X/r^2\), rather than a property of an already excised orbit. The ISCO on the timelike circular branch connected to infinity is fixed by
\begin{equation}
 \left[r f_X f''_X+3f_X f'_X-2r(f'_X)^2\right]_{r=r_{\mathrm I,X}}=0.
 \label{eq:seed-isco}
\end{equation}
The radius \(r_{\mathrm I,X}\) denotes this seed ISCO. Its circular branch additionally requires \(f_X>0\), \(f'_X>0\), and \(2f_X-rf'_X>0\). The ISCO condition follows from marginal radial stability of a timelike geodesic at fixed particle energy and angular momentum; it is distinct from stability of the material shell. In the Schwarzschild limit these radii are \(3M\) and \(6M\), respectively. Their dependence on \(\ell/M\) carries the running-coupling correction into both the null barrier and a disk truncated at the ISCO, as in the disk construction of Ref.~\cite{Reboucas2026ExponentialTSW}.

\section{Symmetric thin-shell construction}\label{sec:junction}

\subsection{Embedding and induced geometry}\label{subsec:embedding}

For a chosen prescription \(X\), take two identical copies of its region \(r\geq R(\tau)\) and identify their spherical boundaries. Here \(R(\tau)>r_{X,+}\) is the throat radius and \(\tau\) is shell proper time. The signs \(+\) and \(-\) now label the two copies rather than the two roots of \cref{eq:horizon-C}. Identical \(M\), \(\ell\), and \(X\) on both sides define a reflection-symmetric wormhole. Its shell \(\Sigma\) has intrinsic coordinates \(\xi^i=(\tau,\theta,\phi)\), where Latin indices \(i,j\) run over these three coordinates. Its embedding in each copy is \(x_\pm^\mu(\xi^i)=(t_\pm(\tau),R(\tau),\theta,\phi)\).

For the junction calculation write \(f=f_X\) for the selected lapse and \(g_{\mu\nu}\) for the metric components of \cref{eq:metric}. A dot denotes \(d/d\tau\); a prime on \(f\) denotes a derivative with respect to its radial argument. Future-directed proper-time normalization gives \(\dot t_\pm=\sqrt{f(R)+\dot R^2}/f(R)\). The induced metric is the pullback
\begin{equation}
 h_{ij}
 =g_{\mu\nu}\,
   \frac{\partial x^\mu_\pm}{\partial\xi^i}
   \frac{\partial x^\nu_\pm}{\partial\xi^j}.
 \label{eq:pullback}
\end{equation}
The tensor \(h_{ij}\) measures intervals tangent to \(\Sigma\) and agrees when evaluated from either copy. Substitution of the normalized embedding gives
\begin{equation}
 ds_\Sigma^2=-d\tau^2
             +R(\tau)^2\left(d\theta^2+\sin^2\theta\,d\phi^2\right).
 \label{eq:induced}
\end{equation}
The quantity \(ds_\Sigma^2\) is the intrinsic shell interval. The area \(4\pi r^2\) increases away from the identified boundary on either side, so \(R\) is a minimum of areal radius across the throat. The retained static exteriors satisfy \(f(R)>0\); the dynamical square roots require \(f(R)+\dot R^2>0\).

Write the boundary locally as \(\Phi_\pm=0\), with \(\Phi_\pm(t_\pm,r)=r-R(\tau(t_\pm))\). A common orientation across the glued manifold, from the minus copy toward the plus copy, gives
\begin{equation}
 n_\mu^\pm
 =\left.\frac{\pm\partial_\mu\Phi_\pm}
 {\sqrt{g^{\alpha\beta}\partial_\alpha\Phi_\pm\partial_\beta\Phi_\pm}}
 \right|_\Sigma
 =\pm\left(-\dot R,\frac{\sqrt{f(R)+\dot R^2}}{f(R)},0,0\right).
 \label{eq:normal}
\end{equation}
The covectors \(n_\mu^\pm\) are the unit spacelike normals in the respective charts, \(g^{\alpha\beta}\) is the inverse metric, and \(\partial_\mu\) is a coordinate derivative. The indices \(\alpha,\beta\) are contracted spacetime indices. These normals obey \(n_\mu^\pm n_\pm^\mu=1\) and are orthogonal to the timelike shell tangent. Opposite radial signs implement the wormhole branch.

\subsection{Extrinsic curvature and surface stresses}\label{subsec:stresses}

We use the extrinsic curvature convention
\begin{equation}
 K^\pm_{ij}
 =-n_\mu^\pm\left(
 \frac{\partial^2x_\pm^\mu}{\partial\xi^i\partial\xi^j}
 +\Gamma^\mu_{\alpha\beta}
 \frac{\partial x_\pm^\alpha}{\partial\xi^i}
 \frac{\partial x_\pm^\beta}{\partial\xi^j}\right).
 \label{eq:extrinsic-definition}
\end{equation}
The second fundamental form \(K^\pm_{ij}\) is evaluated with the bulk Levi-Civita connection \(\Gamma^\mu_{\alpha\beta}\). Its components in the orthonormal shell frame are
\begin{equation}
 K^{\hat\tau}{}_{\hat\tau}\big|_\pm=\pm\frac{\ddot R+f'(R)/2}{\sqrt{f(R)+\dot R^2}},\qquad K^{\hat\theta}{}_{\hat\theta}\big|_\pm=\pm\frac{\sqrt{f(R)+\dot R^2}}{R}.
 \label{eq:extrinsic-components}
\end{equation}
Hats identify the physical orthonormal directions \((\hat\tau,\hat\theta,\hat\phi)\). Spherical symmetry gives \(K^{\hat\phi}{}_{\hat\phi}\big|_\pm=K^{\hat\theta}{}_{\hat\theta}\big|_\pm\). Square brackets denote a jump: the value on the plus copy minus the value on the minus copy. The Einstein--Israel equation in this same frame is
\begin{equation}
 S^{\hat i}{}_{\hat j}
 =-\frac{1}{8\pi}
 \left([K^{\hat i}{}_{\hat j}]
       -\delta^{\hat i}{}_{\hat j}[K]\right).
 \label{eq:israel}
\end{equation}
The tensor \(S^{\hat i}{}_{\hat j}\) is the shell stress tensor, \(\delta^{\hat i}{}_{\hat j}\) is the identity tensor, and \([K]=[K^{\hat i}{}_{\hat i}]\) is the trace jump in that frame~\cite{Israel1966,PoissonVisser1995}. Its isotropic surface-fluid interpretation is
\begin{equation}
 S^{\hat i}{}_{\hat j}=\operatorname{diag}(-\sigma,p,p).
 \label{eq:surface-tensor}
\end{equation}
Here \(\sigma\) is surface energy density, \(p\) is tangential surface pressure, and \(\operatorname{diag}\) denotes the diagonal mixed tensor. Both \(\sigma\) and \(p\) have dimensions of inverse length. The temporal component of the junction equation yields
\begin{equation}
 \sigma=-\frac{\sqrt{f(R)+\dot R^2}}{2\pi R},
 \label{eq:sigma}
\end{equation}
so the wormhole orientation requires negative surface density. The angular component gives
\begin{equation}
 p=\frac{2\ddot R+f'(R)}{8\pi\sqrt{f(R)+\dot R^2}}
   +\frac{\sqrt{f(R)+\dot R^2}}{4\pi R}.
 \label{eq:pressure}
\end{equation}
The two terms represent temporal and angular extrinsic-curvature contributions. With a prescribed pressure function this is also an acceleration equation,
\begin{equation}
 \ddot R=4\pi p\sqrt{f(R)+\dot R^2}
          -\frac{f'(R)}2-\frac{f(R)+\dot R^2}{R}.
 \label{eq:acceleration}
\end{equation}
It determines the radial acceleration locally together with the density branch fixed by the junction equation.

At a static radius \(R_0>r_{X,+}\), define \(f_0=f(R_0)>0\), \(f'_0=f'(R_0)\), and \(f''_0=f''(R_0)\). Subscript \(0\) means equilibrium, with \(\dot R=\ddot R=0\). Its stresses are
\begin{equation}
 \sigma_0=-\frac{\sqrt{f_0}}{2\pi R_0},
 \qquad
 p_0=\frac{2f_0+R_0f'_0}{8\pi R_0\sqrt{f_0}}.
 \label{eq:static-stresses}
\end{equation}
These geometric values must satisfy the chosen equation of state; arbitrary material parameters need not support the selected radius. They give
\begin{equation}
 \sigma_0+p_0
 =\frac{R_0f'_0-2f_0}{8\pi R_0\sqrt{f_0}}.
 \label{eq:tangential-null}
\end{equation}
The intrinsic NEC, tested on null directions tangent to the shell, requires \(\sigma_0+p_0\geq0\). The intrinsic weak and dominant energy conditions fail because \(\sigma_0<0\). Even if the tangent null inequality holds, the distributional four-dimensional NEC fails for suitable transverse null directions: a normally crossing null ray samples the negative \(\sigma_0\). A tangent test therefore does not remove the exotic matter required by this construction.

\begin{figure}[htp!]
    \centering
    \includegraphics[width=0.98\linewidth]{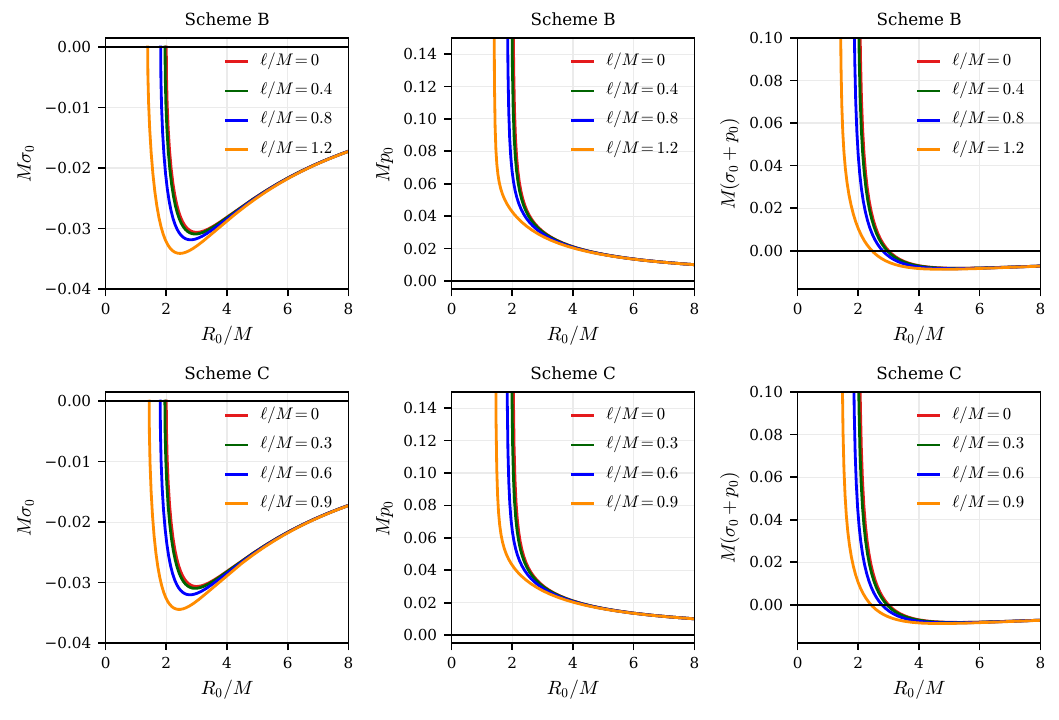}
    \caption{Static junction stresses as functions of the dimensionless throat radius $R_0/M$, with $M=1$. The upper and lower rows correspond to schemes B and C; the columns show the surface density $M\sigma_0$, surface pressure $Mp_0$, and tangential NEC combination $M(\sigma_0+p_0)$. The four equally spaced deformation values are $\ell/M=0,\,0.40,\,0.80,\,1.20$ for B and $\ell/M=0,\,0.30,\,0.60,\,0.90$ for C. Both upper values lie below the respective extremal endpoints in Fig.~\ref{fig:asg-extremal-outer-horizons}. Each curve is restricted to $R_0>r_{X,+}$; the black horizontal line marks zero. Each point gives the junction stresses required at that static radius. These scheme-specific ranges are distinct from the common nonextremal domain used for comparisons at equal deformation.}
    \label{fig:asg-static-stresses}
\end{figure}

The negative density and positive pressure in Fig.~\ref{fig:asg-static-stresses} are the junction stresses required by \cref{eq:static-stresses}. The quantum suppression of the Newton coupling changes the lapse and its radial gradient, thereby changing both stresses without specifying a quantum equation of state for the shell. Within the displayed common exterior of each row, increasing \(\ell/M\) makes the density more negative and lowers the pressure. The density minimum also moves inward, consistently with the horizon displacement in Fig.~\ref{fig:asg-extremal-outer-horizons}. The \(\ell/M=0\) curves coincide between the prescriptions and represent the Schwarzschild thin-shell configuration. At larger radii, their separation weakens according to \cref{eq:asymptotic}.

The tangential NEC is satisfied just outside the outer seed horizon and changes sign at a finite exterior radius. Solving \cref{eq:tangential-null} places this zero at \(R_0/M=3\) for Schwarzschild and approximately \(2.45\) and \(2.42\) for the last B and C curves, respectively. Differentiating the density in \cref{eq:static-stresses} shows that the same zero marks its minimum. The positive pressure therefore compensates the negative density only over a restricted radial interval; the density remains negative throughout the plotted exteriors.

The upper C sample, \(\ell/M=0.90\), keeps its two seed horizons separated by approximately \(0.872M\), avoiding the compressed radial transition associated with their near coalescence. It is a selected subextremal sample within the full family of Fig.~\ref{fig:asg-extremal-outer-horizons}. Every displayed horizon is simple, with \(f'_X(r_{X,+})>0\). The square-root factors in \cref{eq:static-stresses,eq:tangential-null} imply \(\sigma_0\propto-\sqrt{R_0-r_{X,+}}\), whereas \(p_0\) and \(\sigma_0+p_0\) grow as \((R_0-r_{X,+})^{-1/2}\) when approached from outside. The smoother displayed transition thus preserves the divergent stress limit at the excluded horizon; pressure and NEC leave the upper plotting boundary before reaching it.

\subsection{Flux and surface conservation}\label{subsec:conservation}

In the adopted Einstein effective description, the reciprocal temporal and radial metric coefficients imply
\begin{equation}
 T^t{}_t=T^r{}_r,
 \qquad \rho_{\rm b}+p_{\rm b}^{\,r}=0 .
 \label{eq:bulk-radial}
\end{equation}
Here \(T^\mu{}_\nu\) is the effective bulk stress tensor, \(\rho_{\rm b}=-T^t{}_t\) its static-frame density, and \(p_{\rm b}^{\,r}=T^r{}_r\) its radial pressure. The two relations express the same radial-vacuum property. The energy flux \(T_{\mu\nu}u^\mu n^\nu\) consequently vanishes on either side, where \(u^\mu=\partial x^\mu/\partial\tau\) is the unit timelike shell tangent. For a radially moving shell the flux is proportional to \(\rho_{\rm b}+p_{\rm b}^{\,r}\), so it also vanishes.

The shell balance law thus has no bulk flux term:
\begin{equation}
 \frac{d}{d\tau}\bigl(\sigma\mathcal A\bigr)
 +p\,\frac{d\mathcal A}{d\tau}=0 .
 \label{eq:conservation}
\end{equation}
The area is \(\mathcal A=4\pi R^2\). This is mechanical surface energy balance, without assumptions about temperature or entropy. For a differentiable barotropic equation of state \(p=p(\sigma)\), its radial form is
\begin{equation}
 \frac{d\sigma}{dR}=-\frac{2}{R}(\sigma+p).
 \label{eq:density-gradient}
\end{equation}
It defines a local density branch through \(\sigma(R_0)=\sigma_0\). The equation extends continuously to equilibrium; division by \(\dot R\) away from a turning point is not an additional assumption about a moving equilibrium. On an interval with \(\sigma+p\ne0\), the implicit integral is
\begin{equation}
 \ln\!\left(\frac{R}{R_0}\right)
 =-\frac12\int_{\sigma_0}^{\sigma(R)}
              \frac{d\bar\sigma}{\bar\sigma+p(\bar\sigma)}.
 \label{eq:implicit-conservation}
\end{equation}
The integration variable \(\bar\sigma\) is a dummy surface density. At \(\sigma+p=0\), one uses \cref{eq:density-gradient} rather than dividing by that quantity in this integral.

\section{Radial equilibrium and linear stability}\label{sec:stability}

The density junction equation becomes a one-dimensional radial constraint,
\begin{equation}
 \dot R^2+\mathcal V(R)=0,
 \qquad
 \mathcal V(R)=f(R)-[2\pi R\sigma(R)]^2.
 \label{eq:potential}
\end{equation}
The dimensionless \(\mathcal V\) is the shell effective potential. Its density branch solves \cref{eq:density-gradient} for a fixed equation of state. It is not the family of static geometric densities evaluated at successively different radii. Squaring \cref{eq:sigma} retains the wormhole branch only when \(\sigma<0\) is imposed separately.

For a static configuration satisfying its equation of state and \cref{eq:static-stresses},
\begin{equation}
 \mathcal V(R_0)=0,
 \qquad \mathcal V'(R_0)=0 .
 \label{eq:equilibrium}
\end{equation}
Primes on \(\mathcal V\) denote derivatives along its radial density branch. The first equality follows from \(\sigma_0\), and the second uses \(p_0\) and conservation. They do not impose a radius-dependent parameter adjustment during a perturbation.

The local pressure response is
\begin{equation}
 \zeta(\sigma)=\frac{dp}{d\sigma},
 \qquad \zeta_0=\left.\frac{dp}{d\sigma}\right|_{\sigma_0}.
 \label{eq:response}
\end{equation}
The dimensionless \(\zeta\) is differentiated at fixed constitutive parameters. Its interpretation as a causal microscopic sound speed requires a separate material theory; \(0\leq\zeta_0\leq1\) is not imposed on the exotic shell without that additional assumption.

Twice differentiating \cref{eq:potential} with \cref{eq:density-gradient} gives
\begin{equation}
 \mathcal V''(R)
 =f''(R)-8\pi^2\left[
 (\sigma+2p)^2
 +2\sigma(\sigma+p)(1+2\zeta)\right].
 \label{eq:potential-second-general}
\end{equation}
Density, pressure, and response are evaluated on the same dynamical branch. Substituting the static stresses gives
\begin{equation}
 \mathcal V''_0
 =f''_0-\frac{(f'_0)^2}{2f_0}
 -\frac{2f_0-R_0f'_0}{R_0^2}(1+2\zeta_0).
 \label{eq:potential-second-static}
\end{equation}
Here \(\mathcal V''_0=\mathcal V''(R_0)\) has dimensions of inverse length squared. This expression applies to either prescribed geometry without expanded metric derivatives.

Write \(R(\tau)=R_0+\eta(\tau)\), with \(\eta\) a small displacement of dimensions of length. Independently linearizing \cref{eq:acceleration}, using the first-order pressure and density changes \(\delta p\) and \(\delta\sigma\), related by \(\delta p=\zeta_0\delta\sigma\), and \cref{eq:density-gradient}, yields
\begin{equation}
 \ddot\eta+\frac{\mathcal V''_0}{2}\eta=0 .
 \label{eq:linear-motion}
\end{equation}
All background and constitutive parameters are fixed. Thus \(\mathcal V''_0>0\) gives linear radial restoring behavior, \(\mathcal V''_0<0\) an exponentially growing mode, and \(\mathcal V''_0=0\) a marginal case requiring higher-order analysis. In the stable case the proper-time frequency is \(\Omega_0=\sqrt{\mathcal V''_0/2}\), with dimensions of inverse length. On the unchanged zero-energy constraint, an isolated strict minimum with \(\mathcal V(R_0)=0\) has no neighboring finite-amplitude allowed trajectory with identical integration constants. The standard linear criterion concerns restoring acceleration under admissible small perturbations; it does not establish a finite-amplitude periodic orbit on that same zero-energy surface~\cite{PoissonVisser1995}.

To separate geometry from material response, define
\begin{equation}
 D_0=2f_0-R_0f'_0 .
 \label{eq:D}
\end{equation}
The dimensionless \(D_0\) multiplies the response in \cref{eq:potential-second-static} and has sign opposite to \(\sigma_0+p_0\). When \(D_0\ne0\), the marginal response is
\begin{equation}
 \zeta_{\rm crit}
 =-\frac12+\frac{R_0^2}{2D_0}
       \left[f''_0-\frac{(f'_0)^2}{2f_0}\right].
 \label{eq:critical-response}
\end{equation}
The subscript \({\rm crit}\) denotes zero potential curvature. Stability requires \(\zeta_0<\zeta_{\rm crit}\) for \(D_0>0\), and \(\zeta_0>\zeta_{\rm crit}\) for \(D_0<0\). At \(D_0=0\), this division is invalid and \cref{eq:potential-second-static} applies directly. If \(R_0\) coincides with an unstable circular null orbit of the seed, then \((f/r^2)'=0\) and \(\mathcal V''_0=R_0^2(f/r^2)''|_{R_0}<0\), independently of finite barotropic response. This statement concerns an orbit retained from the seed exterior.

In the Schwarzschild limit the marginal response is
\begin{equation}
 \zeta_{{\rm crit},\mathrm S}
 =-\frac{R_0^2-3MR_0+3M^2}
 {2(R_0-2M)(R_0-3M)}, \qquad R_0>2M,\quad R_0\ne3M .
 \label{eq:schwarzschild-critical}
\end{equation}
The additional \(\mathrm S\) label identifies the classical control. The sign of \(D_0\) changes at \(3M\); stability requires the upper-response inequality inside that radius and the lower-response inequality outside it. At \(R_0=3M\), the curvature is \(-2/(9M^2)\), so no finite-response barotrope stabilizes the throat. This recovers the classical shell criterion~\cite{PoissonVisser1995}.

\section{Cosmic Chaplygin surface equations of state}\label{sec:eos}

The EOS fixes pressure as a function of density on a neighborhood of equilibrium. The cosmic Chaplygin forms originated as cosmological volume-fluid laws~\cite{GonzalezDiaz2003,Pourhassan2013}. Their use on a two-dimensional shell is a phenomenological constitutive assumption. Cosmological constraints and homogeneous-fluid stability results do not establish stability of this shell.

Introduce a positive, fixed reference density \(\sigma_\star\), of dimensions of inverse length, and define
\begin{equation}
 s=\frac{\sigma}{\sigma_\star}.
 \label{eq:normalized-density}
\end{equation}
The dimensionless \(s\) is negative on the wormhole branch. The scale \(\sigma_\star\) normalizes the selected EOS and is fixed during perturbations. The choice \(\sigma_\star=1/M\) is possible for fixed \(M\). Setting it equal to a varying function of \(R\) would instead change the pressure response.

For comparison of the two geometries, define \(x=R_0/M\) as the dimensionless radius, \(q=\ell/M\) as the dimensionless deformation, and \(d=M\sigma_\star>0\) as the dimensionless material scale. The sign of \(M^2\mathcal V''_0\) is independent of the length unit. Equal \(x,q,d\) and equal free EOS parameters compare the prescribed geometries at identical scales wherever both exterior conditions hold. Equal \(R_0/r_{X,+}\) instead compares equal relative horizon clearance and generally uses different \(x\). The two choices therefore define different geometric comparisons.

For \(s<0\), use the ordinary real rational power,
\begin{equation}
 s^{m/n}=(-1)^m|s|^{m/n},
 \qquad m,n\in\mathbb Z,\quad n>0\ {\rm odd},\quad \gcd(m,n)=1 .
 \label{eq:real-power}
\end{equation}
Here \(m,n\) are the reduced numerator and positive denominator of the exponent, \(\mathbb Z\) denotes the integers, and \(\gcd\) the greatest common divisor. This convention retains the usual derivative rule on \(s<0\). Replacing \(s\) with its magnitude in the constitutive law, or extending it by a signed power for arbitrary irrational exponents, would define different EOSs.

\subsection{Generalized cosmic Chaplygin gas}\label{subsec:gccg}

The GCCG obeys
\begin{equation}
 p_{\mathrm G}(\sigma)
 =-\sigma_\star s^{-\gamma}
   \left[E+\left(s^{1+\gamma}-E\right)^{-\omega}\right].
 \label{eq:gccg}
\end{equation}
The label \(\mathrm G\) denotes this EOS, \(E\) is a dimensionless pressure parameter, \(\gamma\) the dimensionless Chaplygin exponent, and \(\omega\) the dimensionless cosmic exponent. For this branch, the shell pressure \(p\) equals \(p_{\mathrm G}\). This is the dimensional surface version of the law introduced by Gonz\'alez-D\'iaz~\cite{GonzalezDiaz2003}. Its source amplitude parametrization is
\begin{equation}
 E=\frac{\mathcal B}{1+\omega}-1 .
 \label{eq:E-parameter}
\end{equation}
The dimensionless \(\mathcal B\) can have either sign. We use \(E\) as the constitutive parameter; the inverse conversion is \(\mathcal B=(1+\omega)(E+1)\). The value \(\omega=-1\) is excluded from this amplitude parametrization. At fixed finite \(E\), the formal GCCG pressure at that endpoint simplifies to \(p_{\mathrm G}=-\sigma\), independently of \(E\); it is a separate degenerate law, rather than a tunable cosmic branch.

The original cosmological construction states \(-L_\omega<\omega<0\), where the dimensionless bound \(L_\omega>0\) may exceed unity. It discusses positive Chaplygin exponents and imposes additional exponent relations in particular late-time regimes. It gives no microscopic allowed interval for a negative-density surface layer. We adopt \(0<\gamma\leq1\), with \(\gamma\) rational and of odd reduced denominator according to \cref{eq:real-power}, and \(\omega<0,\omega\ne-1\). No finite lower bound on \(\omega\) follows from the local shell algebra. These shell assumptions define admissible real branches, without importing cosmological observational bounds. The local mathematical domain is broader: rational \(\gamma\) with odd reduced denominator and any real \(\omega\) give real differentiable pressures when \(s<0\) and the secondary base is positive. In particular, \(\gamma=0\) is mathematically regular but excluded from the adopted positive-exponent family. The special amplitude parametrization and its \(\omega=-1\) exception remain separate restrictions. The analytical domain is not restricted to a finite numerical sampling subset.

Reality and differentiability further require
\begin{equation}
 Y=s^{1+\gamma}-E>0 .
 \label{eq:secondary-base}
\end{equation}
The dimensionless \(Y\) is the secondary-power base. Strict positivity allows arbitrary real \(\omega\) and avoids a singular or nondifferentiable endpoint. Both \(s<0\) and \(Y>0\) must hold on an open neighborhood of equilibrium. The parameter \(E\) affects the pressure level and real domain. The exponent \(\gamma\) changes the negative-density prefactor and the density dependence of \(Y\); \(\omega\) changes its secondary-power response. Continuation to negative surface density need not preserve the cosmological pressure sign.

Differentiation at fixed \(E,\gamma,\omega,\sigma_\star\) gives
\begin{equation}
 \zeta_{\mathrm G}(\sigma)
 =-\gamma\frac{p_{\mathrm G}(\sigma)}{\sigma}
   +\omega(1+\gamma)Y^{-1-\omega}.
 \label{eq:gccg-response}
\end{equation}
The response label \(\mathrm G\) specifies the GCCG branch. Its first term depends on the pressure-to-density ratio; the explicitly cosmic term is negative on the adopted domain. Their sum determines the restoring coefficient. At fixed \(\sigma\) there is no explicit radius dependence, so no further partial derivative with respect to \(R\) enters.

The factor \(\sigma_\star\) supplies the dimension of surface pressure; every powered quantity in the brackets is dimensionless. In the limit \(\omega\to0\) with fixed \(\mathcal B\), the pressure in \cref{eq:gccg} becomes \(p_{\mathrm G}=-\mathcal B\sigma_\star s^{-\gamma}\), the generalized Chaplygin surface law. Taking also \(\gamma=1\) gives \(p_{\mathrm G}=-\mathcal B\sigma_\star^2/\sigma\), the ordinary Chaplygin law.

A static throat must satisfy
\begin{equation}
 \frac{p_0}{\sigma_\star}
 =-s_0^{-\gamma}
  \left[E_{\mathrm G,0}
        +\left(s_0^{1+\gamma}-E_{\mathrm G,0}\right)^{-\omega}\right],
 \qquad s_0=\frac{\sigma_0}{\sigma_\star}.
 \label{eq:gccg-static}
\end{equation}
Here \(s_0<0\) is the normalized static density and \(E_{\mathrm G,0}\) is the calibrated \(E\) for this throat. Multiplication by the real \(s_0^\gamma\) turns this into
\begin{equation}
 Y_{\mathrm G,0}-Y_{\mathrm G,0}^{-\omega}
 =s_0^{1+\gamma}+\frac{p_0}{\sigma_\star}s_0^\gamma,
 \qquad Y_{\mathrm G,0}>0 .
 \label{eq:gccg-calibration}
\end{equation}
The positive root \(Y_{\mathrm G,0}\) fixes \(E_{\mathrm G,0}=s_0^{1+\gamma}-Y_{\mathrm G,0}\). For the existence test, denote the dimensionless right-hand side of \cref{eq:gccg-calibration} by \(Q\) and set \(k=-\omega>0\), with \(k\ne1\). For \(0<k<1\), positive roots exist exactly when \(Q\geq-(1-k)k^{k/(1-k)}\), with two roots strictly between that lower bound and zero. For \(k>1\), existence requires \(Q\leq(k-1)k^{-k/(k-1)}\), with two roots strictly between zero and that upper bound. At either bound the roots merge; every other allowed \(Q\) gives one positive root. The zero root is excluded. Merging makes the calibration map nonregular, but the fixed-\(E\) EOS remains differentiable for \(s<0\) and \(Y>0\).

Distinct calibration roots specify different material models. The selected \(E_{\mathrm G,0}\) and its converted amplitude remain fixed throughout a perturbation; recalibration as \(R\) changes would replace the material rather than perturb one shell.

Substituting \cref{eq:gccg-response} at equilibrium into \cref{eq:potential-second-static} gives
\begin{equation}
 \mathcal V''_{\mathrm G,0} =f''_0-\frac{(f'_0)^2}{2f_0}-\frac{D_0}{R_0^2} \left[1-2\gamma\frac{p_0}{\sigma_0} +2\omega(1+\gamma)Y_{\mathrm G,0}^{-1-\omega}\right].
 \label{eq:gccg-stability}
\end{equation}
The GCCG curvature \(\mathcal V''_{\mathrm G,0}\) must be positive for linear radial restoring behavior. This is necessary and sufficient within the stated fixed-background barotropic perturbation problem. Existence, exterior, and real-branch conditions are separate prerequisites.

For the sampled endpoint \(\gamma=1\), the static stress ratio cancels the inverse-lapse contribution in \cref{eq:gccg-stability}, leaving
\begin{equation}
 \mathcal V''_{\mathrm G,0}=f''_0+\frac{f'_0}{R_0}-\frac{4f_0}{R_0^2}
 -\frac{4\omega D_0}{R_0^2}Y_{\mathrm G,0}^{-1-\omega}.
 \label{eq:gccg-unit-exponent-curvature}
\end{equation}
This is a specialization of the same ordinary-power GCCG, with no change of its constitutive function. For a finite positive calibration root, its near-horizon limit is governed by finite lapse derivatives and the cosmic response. The quantum-running geometry can therefore change the sign of that limit through its radial profile, even though the junction law and local perturbation treatment remain those of the effective Einstein description.

The numerical families use \(M=1\), \(d=1\), and the largest positive calibration root. The rational exponents \(\gamma=1/5,3/5,1\) preserve \cref{eq:real-power}; no magnitude substitution or signed extension is used. The five deformations cover the same endpoints as the static-stress families, with one additional sample. Scheme B includes its separate nonextremal extension to \(\ell/M=1.20\); C ends at \(0.90\). Direct comparisons of the prescriptions at equal deformation are restricted to their shared range. The horizontal window \(1.3\leq R_0/M\leq10\) starts just below the smallest sampled seed horizon, retains both sides of every photon sphere, and extends beyond all corresponding ISCO radii. Each plotted point fixes a new equilibrium amplitude, whereas that amplitude stays fixed during the perturbation defining its curvature.

\begin{figure}[htp!]
    \centering
    \includegraphics[width=0.98\linewidth]{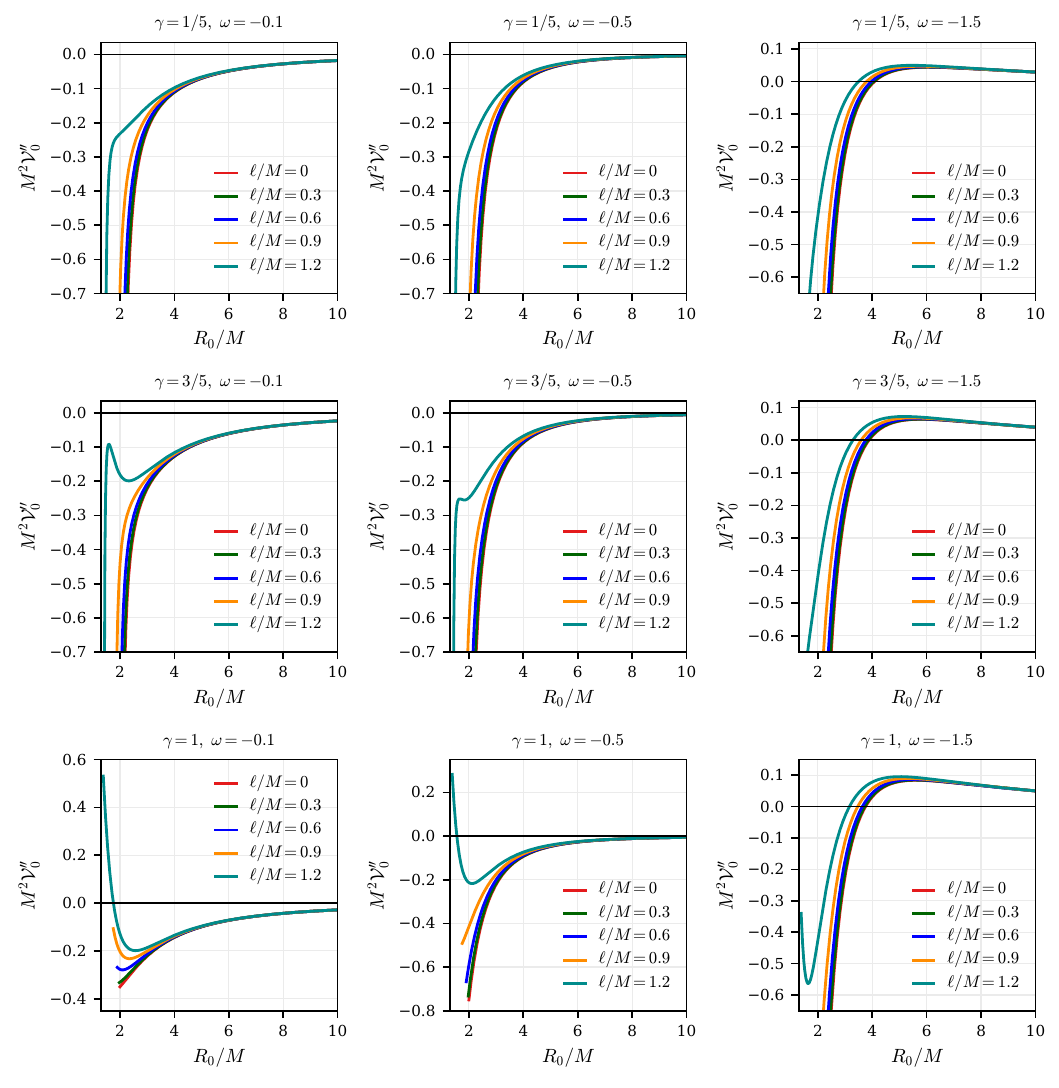}
    \caption{GCCG radial stability curvature $M^2\mathcal V^{\prime\prime}_0$ as a function of $R_0/M$ for scheme B, with $M=1$ and fixed density scale $d=M\sigma_\star=1$. Rows use $\gamma=1/5,\,3/5,\,1$; columns use $\omega=-0.1,\,-0.5,\,-1.5$. The five curves have $\ell/M=0,\,0.30,\,0.60,\,0.90,\,1.20$. Ordinary real rational powers and the largest positive root of Eq.~\eqref{eq:gccg-calibration} are used. Each point is a separately calibrated static shell; its material amplitude is held fixed during the local perturbation. Only the outer static domain and real EOS branch are drawn. Positive and negative curvature denote stable and unstable radial response, separated by the black zero line. All panels use $1.3\leq R_0/M\leq10$; ordinate windows differ as labeled and clip values outside the displayed ranges, including divergent near-horizon tails. Each panel has a vertical legend.}
    \label{fig:asg-gccg-b}
\end{figure}

The curve beginnings in Fig.~\ref{fig:asg-gccg-b} follow the exterior and real-calibration restrictions, rather than a universal lower radius \(2M\). The Schwarzschild horizon is \(2M\), whereas the B horizons move inward as \(\ell/M\) increases, reaching approximately \(1.387M\) in the last curve. The positive-root condition in \cref{eq:gccg-calibration} can further restrict the material branch. Values outside the ordinate windows are clipped, so the first visible point need not be the physical branch endpoint. Each curve is a family of separately calibrated equilibria; its shape is distinct from the displacement dependence of one shell with fixed material coefficients.

The narrow inner maximum is most pronounced for \(\gamma=3/5\), \(\omega=-0.1\), and \(\ell/M=1.20\). It occurs at \(R_0/M\simeq1.591\), where \(M^2\mathcal V''_0\simeq-0.0913\), followed by a minimum near \(2.322\) with curvature approximately \(-0.1989\). Both remain negative. Its origin can be exposed by substituting the static stress ratio into \cref{eq:gccg-stability}: the inverse-lapse contribution becomes \((\gamma-1)(f'_0)^2/(2f_0)\). For \(\gamma<1\), this term is negative and varies rapidly close to a simple horizon wherever the real material branch exists. Its competition with the finite lapse-derivative terms and the cosmic response produces the turnover. At the maximum, the positive secondary base is approximately \(0.996\), away from its excluded zero endpoint. The feature is therefore a finite maximum of the equilibrium-family curvature, without a stability transition or a singularity of the constitutive pressure. The weaker maximum--minimum pair in the \(\omega=-0.5\) panel also stays negative.

The \(\omega=-1.5\) column instead has an outer positive sector for every sampled deformation and exponent. In the Schwarzschild control, the zero crossings occur at \(R_0/M\simeq4.083,3.880,3.750\), from the first to the third row. At \(\ell/M=1.20\), they move to approximately \(3.495,3.290,3.160\), all between \(r_{\mathrm{ph},\mathrm B}/M\simeq2.451\) and \(r_{\mathrm I,\mathrm B}/M\simeq5.071\). These stable cuts excise the seed photon sphere while retaining the ISCO. The broad positive maxima already occur for Schwarzschild: in the \(\gamma=1\) row, its maximum is approximately \(0.08425\) at \(R_0/M\simeq5.626\), compared with \(0.09567\) at \(5.024\) for B at \(\ell/M=1.20\). The deformation raises and moves this maximum inward in the sampled family. The subsequent decrease beyond the ISCO reduces the positive restoring curvature without an additional sign change in the displayed interval. The ISCO is an emitter-orbit scale and does not impose a boundary on shell existence. The maximum need not coincide with the ISCO: the Schwarzschild peak for \(\gamma=1/5\) lies at approximately \(6.071M\), just beyond it. Raising \(\gamma\) from \(1/5\) to \(1\) already increases the classical peak from approximately \(0.04408\) to \(0.08425\) and shifts it inward. This row dependence therefore includes a constitutive effect present before the quantum-running deformation.

For \(\gamma=1\), the inverse-lapse term cancels exactly as in \cref{eq:gccg-unit-exponent-curvature}. At \(\ell/M=1.20\), the \(\omega=-0.1\) and \(-0.5\) curves then have inner stable sectors ending at approximately \(R_0/M=1.758\) and \(1.557\). Both lie above the outer horizon and below the seed photon sphere, which remains in the retained exterior. This change from the Schwarzschild controls depends on the radial profile inherited from quantum running together with the calibrated GCCG response. The shell perturbation itself remains classical in the effective Einstein description. At this same B deformation, the \(\omega=-1.5\) curve first descends to a finite negative minimum, \(M^2\mathcal V''_0\simeq-0.5636\) at \(R_0/M\simeq1.646\), before rising toward its outer stable sector. This dip remains within the unstable region.

At the seed photon sphere, \(D_0=0\), and the negative-curvature result following \cref{eq:critical-response} applies independently of the finite GCCG response. The inner and outer stable sectors therefore cannot join through a stable photon-sphere throat. The local inner maximum and the broad outer maximum describe different radial regimes; only positive curvature supplies linear radial restoring behavior.

\begin{figure}[htp!]
    \centering
    \includegraphics[width=0.98\linewidth]{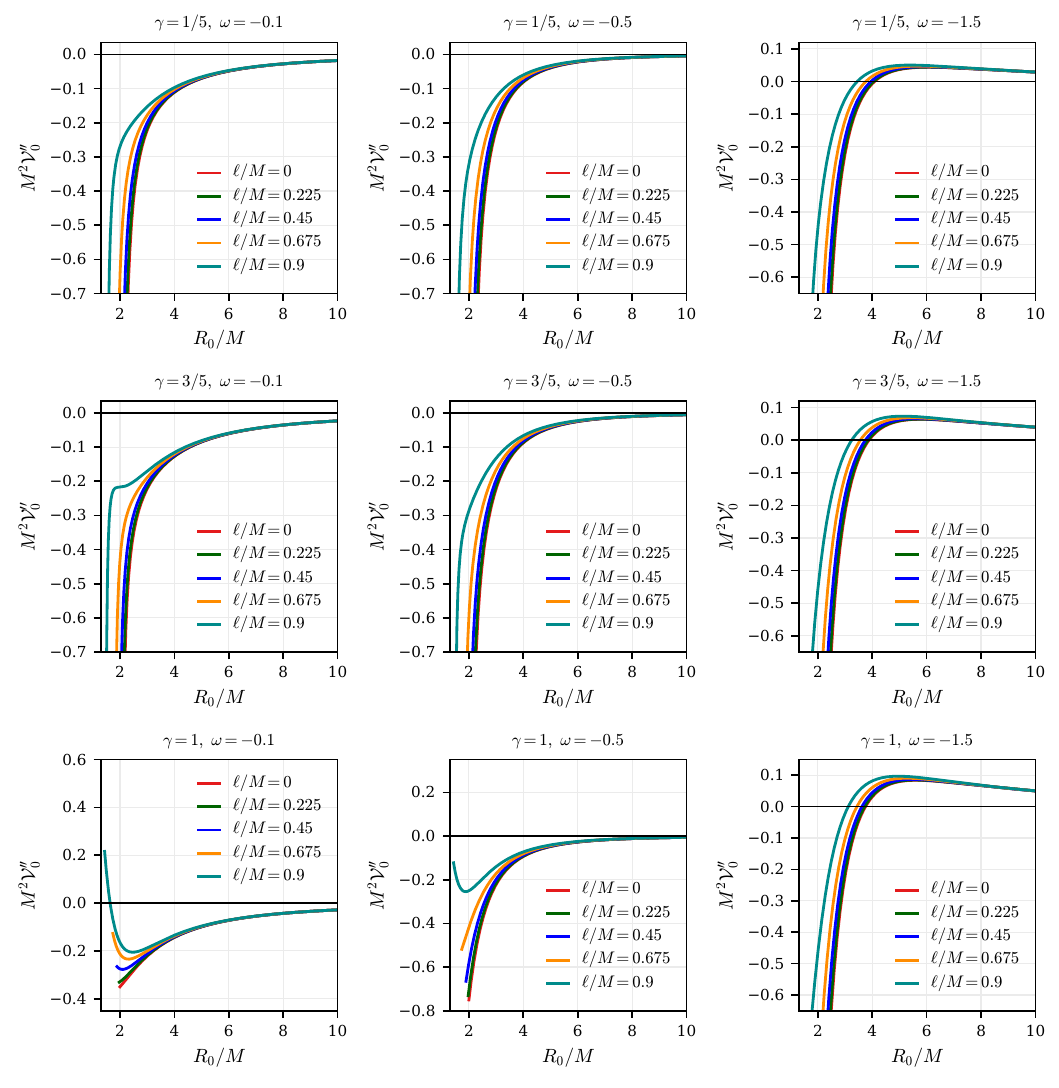}
    \caption{GCCG radial stability curvature $M^2\mathcal V^{\prime\prime}_0$ as a function of $R_0/M$ for scheme C, with $M=1$ and fixed density scale $d=M\sigma_\star=1$. Rows use $\gamma=1/5,\,3/5,\,1$; columns use $\omega=-0.1,\,-0.5,\,-1.5$. The five curves have $\ell/M=0,\,0.225,\,0.450,\,0.675,\,0.900$. Ordinary real rational powers and the largest positive root of Eq.~\eqref{eq:gccg-calibration} are used. Each point is a separately calibrated static shell; its material amplitude is held fixed during the local perturbation. Only the outer static domain and real EOS branch are drawn. Positive and negative curvature denote stable and unstable radial response, separated by the black zero line. All panels use $1.3\leq R_0/M\leq10$; ordinate windows differ as labeled and clip values outside the displayed ranges, including divergent near-horizon tails. Each panel has a vertical legend.}
    \label{fig:asg-gccg-c}
\end{figure}

For scheme C in Fig.~\ref{fig:asg-gccg-c}, the \(\omega=-1.5\) column also supplies stable throats between the seed photon sphere and ISCO for all sampled exponents. At \(\ell/M=0.90\), the crossings are approximately \(R_0/M=3.446,3.245,3.118\), with \(r_{\mathrm{ph},\mathrm C}/M\simeq2.423\) and \(r_{\mathrm I,\mathrm C}/M\simeq4.973\). At the same deformation in B, the crossings are \(3.793,3.592,3.464\), while \(r_{\mathrm{ph},\mathrm B}/M\simeq2.741\) and \(r_{\mathrm I,\mathrm B}/M\simeq5.529\). The larger suppression of \(G_{\mathrm C}/G_0\) at equal normalized cutoff shifts both orbital scales and the restoring boundary farther inward. These comparisons use the same \(M,d,\gamma,\omega\) and root prescription; the material amplitude is determined by each geometry's equilibrium stresses and held fixed only during its local perturbation.

The narrow maximum in B at \(\ell/M=1.20\) is absent from the displayed C families. That B sample lies beyond the common black-hole-seed range, whereas C ends at \(0.90\). Comparing only these last curves would mix a prescription change with a deformation change. At the common value \(\ell/M=0.90\), the \(\gamma=3/5\), \(\omega=-0.1\) curves in both prescriptions have no corresponding inner maximum. The prominent B turnover therefore characterizes its sampled extended deformation range, rather than establishing a universal B/C distinction. Its dependence on the GCCG response also separates this feature from a direct observable of quantum fluctuations.

The broad positive maxima allow a comparison within the shared domain. For \(\gamma=1\) and \(\omega=-1.5\), B at \(\ell/M=0.90\) peaks near \(R_0/M=5.318\) with \(M^2\mathcal V''_0\simeq0.08995\); C peaks near \(4.962\) with curvature approximately \(0.09711\). Both exceed the corresponding Schwarzschild maximum and occur at smaller radii, with the C maximum closer to its own ISCO. A larger peak does not imply a wider interval available for an ISCO-truncated disk. In the same \(\gamma=1\) cases, the width between the stability onset and the ISCO is approximately \(1.856M\) in C and \(2.065M\) in B, compared with \(2.250M\) for Schwarzschild. The quantum-running displacement of the ISCO is therefore as relevant as the displacement of the restoring boundary.

An inner positive branch occurs for C at \(\gamma=1\), \(\omega=-0.1\), and \(\ell/M=0.90\), between the vicinity of \(r_{\mathrm C,+}/M\simeq1.436\) and \(R_0/M\simeq1.637\). The corresponding B horizon is approximately \(1.754M\), so this entire C sector lies outside the common admitted throat domain. Its appearance combines a newly accessible static exterior with the material response. C has no analogous inner positive branch for \(\omega=-0.5\) within the displayed samples. At an admitted common radius, \(R_0/M=4\), with \(\gamma=1/5\) and \(\omega=-1.5\), the Schwarzschild curvature is approximately \(-0.00664\), whereas B and C at \(\ell/M=0.90\) give \(0.01429\) and \(0.03223\). Both deformed throats lie between their seed photon sphere and ISCO. Thus a sign change relative to the classical control also occurs where both effective geometries admit the same radius, although their equilibrium material amplitudes differ.

The enlarged outer window resolves the decay beyond the ISCO and the decreasing separation of the curves. Its continuation follows from \cref{eq:asymptotic,eq:gccg-calibration,eq:gccg-stability}. On the largest-positive-root branch, \(Y_{\mathrm G,0}\to1\) at fixed \(M,\ell,\sigma_\star,\gamma,\omega\) as \(R_0\to\infty\). The leading curvature satisfies \(R_0^2\mathcal V''_{\mathrm G,0}\to-2(1+\gamma)(1+2\omega)\). For the special column \(\omega=-1/2\), that coefficient vanishes, and the next order gives \(R_0^3\mathcal V''_{\mathrm G,0}\to-2(2+\gamma)M\). These limits explain the negative approach to zero in the first two columns, the faster decay in the middle column, and the positive approach in the last. They are shared by B, C and Schwarzschild at the stated material normalization; their subleading geometry dependence produces the finite-radius separations. The plotted interval shows the onset of this decay and does not imply exact equality or complete attainment of the asymptotic limit at \(10M\).

The classical controls, shifted maxima, admitted radial domains and distinct decay rates separate the material response from the effects inherited from quantum gravitational coarse graining. The comparisons concern the stated B/C flow prescriptions and radial scale identification within the IR-inspired background; they do not determine a scheme-independent quantum-gravity prediction or a microscopic state of the shell. The stable inner and outer sectors provide candidates for retaining or excising the seed photon sphere in the optical construction of Ref.~\cite{Reboucas2026ExponentialTSW}. With observer-side emission truncated at the retained ISCO, radial stability supports either choice of cut. Its actual image still requires null trajectories, emitter motion, emissivity and inclination to be calculated.

\subsection{Modified cosmic Chaplygin gas}\label{subsec:mccg}

The MCCG adds a linear density term~\cite{Pourhassan2013}:
\begin{equation}
 p_{\mathrm M}(\sigma)
 =\beta\sigma-\sigma_\star s^{-\gamma}
   \left[E+\left(s^{1+\gamma}-E\right)^{-\omega}\right].
 \label{eq:mccg}
\end{equation}
The label \(\mathrm M\) denotes this EOS and \(\beta\) is its dimensionless linear pressure coefficient. The parameters \(E,\gamma,\omega,\sigma_\star\) have the meanings attached to the full GCCG law, with conversion \cref{eq:E-parameter}. The coefficient \(\beta\) is distinct from the area \(\mathcal A\).

The primary cosmological presentation uses a positive linear coefficient and the cosmic correction, and discusses positive Chaplygin exponents; its viscous completion also introduces dissipative terms~\cite{Pourhassan2013}. Here we adopt only the nondissipative barotropic pressure. No bulk or surface viscosity and no cosmological evolution equation are transferred to the shell. The adopted interval is \(\beta\geq0\), including the \(\beta=0\) GCCG boundary, with the same \(0<\gamma\leq1\) odd-denominator rational exponents and \(\omega<0,\omega\ne-1\). The local derivative exists for other real \(\beta\), but these lie outside this comparison. The cosmological presentation supplies no surface-matter bound forcing \(\beta\leq1\). The analytical relations retain this domain; the numerical examples use a specified subset.

The conditions \(s<0\) and \(Y>0\) from \cref{eq:secondary-base} still apply, now with the MCCG value of \(E\). At fixed density, positive \(\beta\) contributes negative pressure because \(\sigma<0\). The full pressure response is
\begin{equation}
 \zeta_{\mathrm M}(\sigma)
 =\beta(1+\gamma)-\gamma\frac{p_{\mathrm M}(\sigma)}{\sigma}
   +\omega(1+\gamma)Y^{-1-\omega}.
 \label{eq:mccg-response}
\end{equation}
The dimensionless \(\zeta_{\mathrm M}\) is taken at fixed constitutive parameters. The coefficient \(\beta(1+\gamma)\) includes the linear term recovered when the nonlinear pressure is expressed in terms of the total \(p_{\mathrm M}\); replacing it by \(\beta\) in this form would omit that contribution. As in GCCG, there is no explicit radius dependence. At the excluded endpoint \(\omega=-1\), the formal fixed-\(E\) pressure would be \(p_{\mathrm M}=(\beta-1)\sigma\), again losing the calibration freedom in \(E\).

The scale \(\sigma_\star\) ensures the dimension of surface pressure also in this modified law. At \(\beta\to0\), \cref{eq:mccg,eq:mccg-response} reduce to GCCG. At \(\omega\to0\), with \(\mathcal B\) fixed, the pressure becomes \(p_{\mathrm M}=\beta\sigma-\mathcal B\sigma_\star s^{-\gamma}\), the modified Chaplygin surface law. Taking also \(\gamma=1\) gives its linear-plus-inverse-density form.

The static MCCG condition is
\begin{equation}
 \frac{p_0-\beta\sigma_0}{\sigma_\star}
 =-s_0^{-\gamma}
 \left[E_{\mathrm M,0}
       +\left(s_0^{1+\gamma}-E_{\mathrm M,0}\right)^{-\omega}\right].
 \label{eq:mccg-static}
\end{equation}
The dimensionless \(E_{\mathrm M,0}\) is calibrated at fixed \(\beta,\gamma,\omega,\sigma_\star\) for the selected throat. Its root equation is
\begin{equation}
 Y_{\mathrm M,0}-Y_{\mathrm M,0}^{-\omega}
 =s_0^{1+\gamma}
  +\frac{p_0-\beta\sigma_0}{\sigma_\star}s_0^\gamma,
 \qquad Y_{\mathrm M,0}>0 .
 \label{eq:mccg-calibration}
\end{equation}
The selected root \(Y_{\mathrm M,0}\) fixes \(E_{\mathrm M,0}=s_0^{1+\gamma}-Y_{\mathrm M,0}\). This generally differs from the GCCG root at the same geometric throat because the nonlinear pressure required for equilibrium has changed. The same positive-root existence test applies, with \(Q\) now denoting the right-hand side of \cref{eq:mccg-calibration}. Calibration is performed once; its result is fixed during a perturbation.

The potential curvature is consequently
\begin{equation}
 \mathcal V''_{\mathrm M,0} =f''_0-\frac{(f'_0)^2}{2f_0}-\frac{D_0}{R_0^2} \left[1+2\beta(1+\gamma)-2\gamma\frac{p_0}{\sigma_0} +2\omega(1+\gamma)Y_{\mathrm M,0}^{-1-\omega}\right].
 \label{eq:mccg-stability}
\end{equation}
The MCCG curvature \(\mathcal V''_{\mathrm M,0}\) has the same sign interpretation as \cref{eq:linear-motion}. At fixed geometry, comparing the two EOSs requires both the explicit linear contribution and the change of calibration root. The sign of \(\beta\) alone does not establish which EOS has a larger stable region.

The numerical MCCG comparison keeps the normalization, real-power convention, largest-positive-root prescription and five deformation values used for GCCG, with \(M=d=1\). The cosmic exponent is fixed at \(\omega=-1/2\), while the columns vary \(\beta\). For this exponent, \cref{eq:mccg-calibration} has the explicit largest root \(Y_{\mathrm M,0}=\tfrac14[1+\sqrt{1+4Q}]^2\), with \(Q\geq-1/4\). The merging point has \(Y_{\mathrm M,0}=1/4>0\); for \(-1/4<Q<0\), the smaller positive root specifies a different material amplitude and is not used. This algebraic specialization leaves the original signed-density pressure intact. As for GCCG, the selected amplitude is fixed during each perturbation and recalibrated between the equilibria forming a curve.

The explicit linear term changes the curvature by \(-2\beta(1+\gamma)D_0/R_0^2\) when the calibration root is held fixed. Its sign favors restoring response below the seed photon sphere, where \(D_0<0\), and opposes it above that orbit. In the calibrated MCCG/GCCG families, the change of \(Y_{\mathrm M,0}\) contributes alongside the explicit linear term, so that term alone does not imply a monotonic ordering. The geometry still enters through the quantum-running lapse and its derivatives; \(\beta\) describes the phenomenological surface material.

\begin{figure}[htp!]
    \centering
    \includegraphics[width=0.98\linewidth]{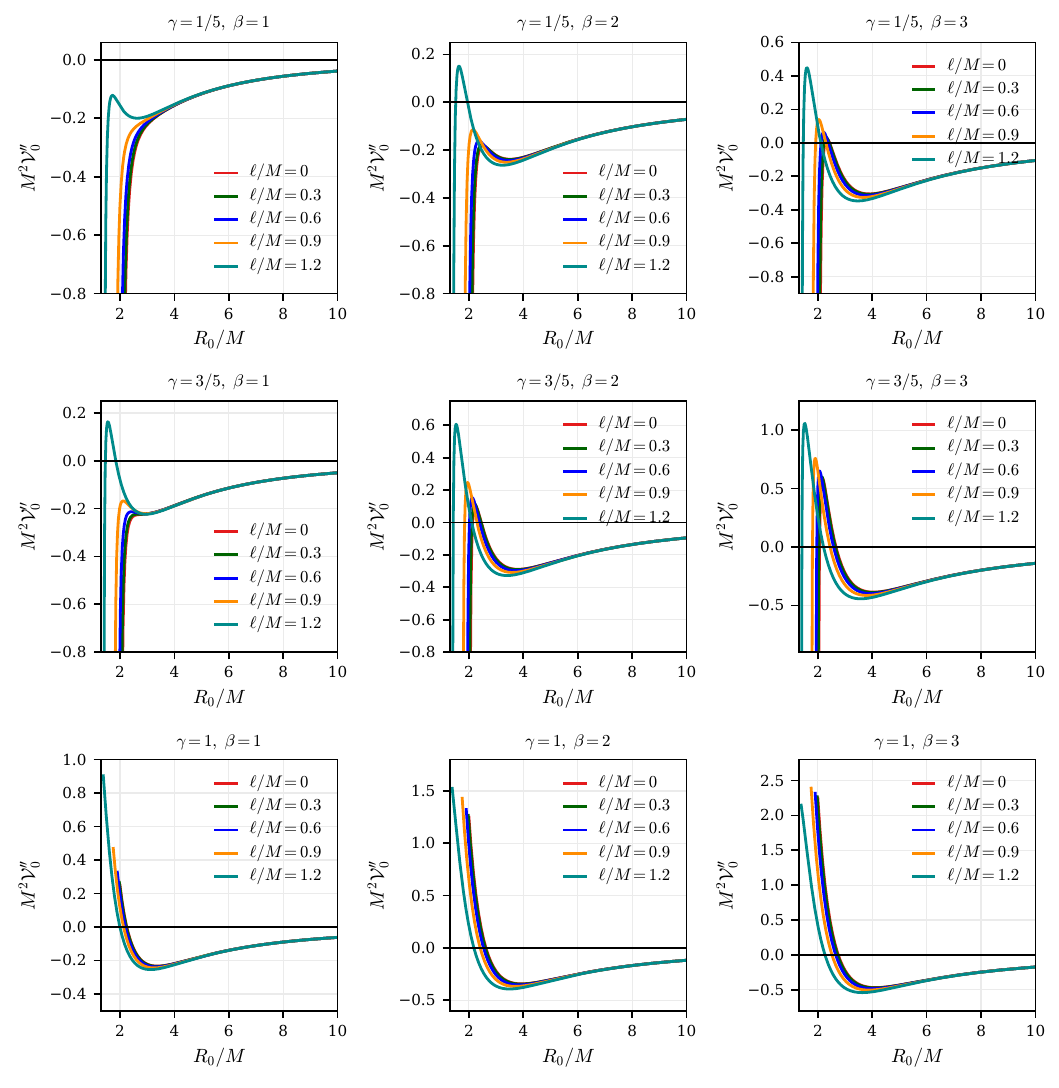}
    \caption{MCCG radial stability curvature $M^2\mathcal{V}^{\prime\prime}_0$ as a function of $R_0/M$ for scheme B, with $M=1$, $d=M\sigma_\star=1$, and $\omega=-0.5$. Rows use $\gamma=1/5,\,3/5,\,1$; columns use $\beta=1,\,2,\,3$. The five curves have $\ell/M=0,\,0.30,\,0.60,\,0.90,\,1.20$. Ordinary real rational powers and the largest positive root of Eq.~\eqref{eq:mccg-calibration} are used. Each point is a separately calibrated equilibrium; its material amplitude is fixed during the local perturbation. Only the outer static region with a real positive calibration root is drawn. Positive and negative curvature denote stable and unstable radial response; the black line marks zero. All panels use $1.3\leq R_0/M\leq10$; ordinate limits clip values outside the displayed windows. Each panel has a compact vertical legend.}
    \label{fig:asg-mccg-b}
\end{figure}
In Fig.~\ref{fig:asg-mccg-b}, the linear coefficient turns some of the inner maxima into positive restoring sectors. The distinction between a peak and stability is visible in the first row: at \(\ell/M=1.20\), \(\beta=1\) gives a maximum \(M^2\mathcal V''_{\mathrm M,0}\simeq-0.1214\) near \(R_0/M=1.726\), whereas \(\beta=2\) gives approximately \(0.1508\) near \(1.636\). The latter has a stable interval \(1.514\lesssim R_0/M\lesssim1.943\). For \(\gamma=3/5\), the same B deformation already gives a positive maximum at \(\beta=1\), approximately \(0.1635\) near \(1.557\). The rise, turnover and subsequent negative trough reflect the calibrated pressure response combined with the radial geometry. They are finite features of an equilibrium family, rather than a time evolution or a direct fluctuation spectrum of the quantum gravitational field.

Inner restoring sectors are not exclusive to the deformed seeds. The Schwarzschild controls have \(2.260\lesssim R_0/M\lesssim2.432\) for \(\gamma=1/5,\beta=3\), and \(2.138\lesssim R_0/M\lesssim2.455\) for \(\gamma=3/5,\beta=2\). The negative curves and near-horizon suppression in other panels therefore cannot be attributed to ASG alone. The inverse-lapse contribution \((\gamma-1)(f'_0)^2/(2f_0)\), identified for GCCG, also occurs after rearranging \cref{eq:mccg-stability}. It is negative for the first two rows and cancels at \(\gamma=1\). In addition to the exterior restriction and ordinate clipping, the positive-root condition can terminate a calibrated family before the horizon is reached. None of these boundaries is a stable throat at the horizon.

For \(\gamma=1\), the curvature has a finite positive outer-horizon limit in all displayed MCCG samples. The Schwarzschild stable sectors extend from the vicinity of \(2M\) to approximately \(R_0/M=2.249,2.625,2.750\) as \(\beta=1,2,3\). At the common deformation \(\ell/M=0.90\), their B upper boundaries move to \(2.131,2.414,2.519\), while the outer horizon moves to \(1.754M\). The inward shift of an upper crossing and the expansion of the admitted exterior are distinct effects. For the additional B sample \(\ell/M=1.20\), the crossings become \(1.998,2.183,2.263\). These last values characterize B's extended deformation range and do not alone isolate a B/C prescription difference.

\begin{figure}[htp!]
    \centering
    \includegraphics[width=0.98\linewidth]{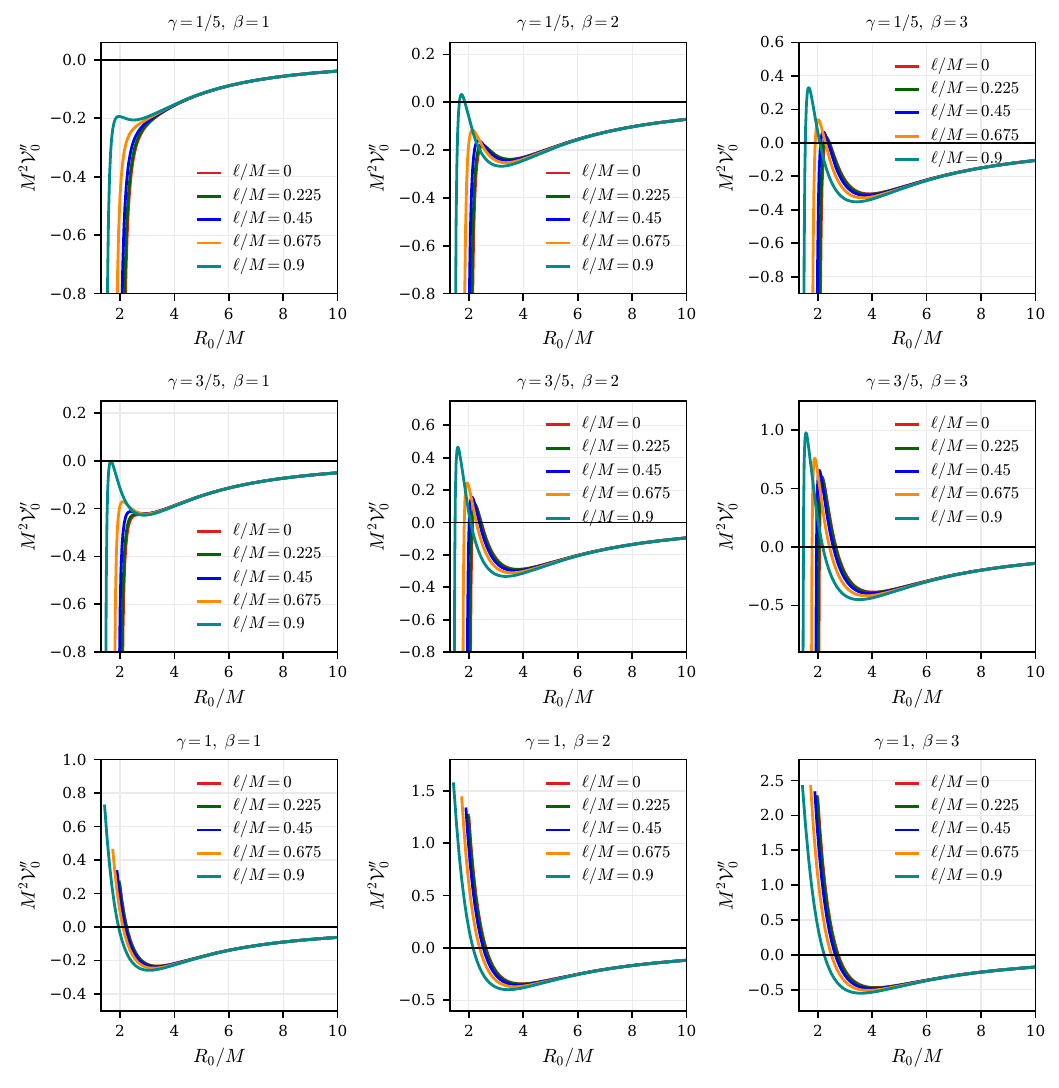}
    \caption{MCCG radial stability curvature $M^2\mathcal{V}^{\prime\prime}_0$ as a function of $R_0/M$ for scheme C, with $M=1$, $d=M\sigma_\star=1$, and $\omega=-0.5$. Rows use $\gamma=1/5,\,3/5,\,1$; columns use $\beta=1,\,2,\,3$. The five curves have $\ell/M=0,\,0.225,\,0.450,\,0.675,\,0.900$. Ordinary real rational powers and the largest positive root of Eq.~\eqref{eq:mccg-calibration} are used. Each point is a separately calibrated equilibrium; its material amplitude is fixed during the local perturbation. Only the outer static region with a real positive calibration root is drawn. Positive and negative curvature denote stable and unstable radial response; the black line marks zero. All panels use $1.3\leq R_0/M\leq10$; ordinate limits clip values outside the displayed windows. Each panel has a compact vertical legend.}
    \label{fig:asg-mccg-c}
\end{figure}
Scheme C in Fig.~\ref{fig:asg-mccg-c} has the same qualitative sequence of inner response, negative trough and distant decay, but can differ in the sign of a narrow maximum. At \(\gamma=3/5,\beta=1,\ell/M=0.90\), its maximum is \(M^2\mathcal V''_{\mathrm M,0}\simeq-0.001626\) near \(R_0/M=1.664\). It approaches the zero line without supplying a stable interval. At the same deformation, B's corresponding maximum is approximately \(-0.1677\); the positive B maximum at \(1.20\) uses a different deformation. In contrast, \(\gamma=1/5,\beta=2\) gives a C restoring interval \(1.650\lesssim R_0/M\lesssim1.852\) at \(\ell/M=0.90\), while the matched B curve stays negative. Part of this interval lies above both outer horizons, although all of it lies below the Schwarzschild horizon, so the classical seed does not admit the same radii.

At equal \(\ell/M=0.90\), with \(\gamma=3/5,\beta=2\), B is stable over approximately \(1.844<R_0/M<2.292\), and C over \(1.490<R_0/M<2.066\). Their positive maxima are approximately \(0.2472\) and \(0.4648\), respectively. C thus moves this interval inward and increases its peak, in addition to lowering the outer-horizon bound. Its stronger suppression of the Newton coupling at equal normalized cutoff changes the static stresses and the material amplitude required for equilibrium. This comparison keeps the free material inputs fixed, but does not identify the independently calibrated amplitudes.

A larger inner response does not mean a larger curvature at every common radius. For \(\gamma=1,\beta=2\), the C stable sector at \(\ell/M=0.90\) ends near \(2.152M\), compared with \(2.414M\) for B and \(2.625M\) for Schwarzschild. At the admitted common radius \(R_0/M=2.2\), their curvatures are approximately \(-0.04570\), \(0.2592\), and \(0.6380\), respectively. The quantum-running deformation can therefore remove a restoring configuration at a fixed radius even while opening a stable sector farther inward. This dependence on radius, material response and calibration precludes a universal claim that either prescription stabilizes every shell more efficiently.

The curves cross into a negative trough before rising toward zero at larger radii. This outer behavior differs from the positive tail of the \(\omega=-1.5\) GCCG column. At the present \(\omega=-1/2\), the largest MCCG root tends to one, and \cref{eq:mccg-response} gives \(\zeta_{\mathrm M,0}\to\beta(1+\gamma)-1/2\). Consequently \(R_0^2\mathcal V''_{\mathrm M,0}\to-4\beta(1+\gamma)\) along the calibrated equilibrium family. The nonzero linear coefficient gives a negative leading tail, while the same cosmic-exponent GCCG column has the faster negative decay discussed above. The shared leading limit explains the decreasing B/C/Schwarzschild separation without implying exact coincidence at \(10M\). Differences inherited from quantum running enter at subleading order there.

All positive MCCG sectors found in these parameter slices lie below their own seed photon sphere; none occupies the interval from that sphere to the ISCO within the examined radial window. In the matched \(\omega=-1/2,\gamma=1\) comparison, \(\beta=1\) extends the B \(\ell/M=1.20\) inner sector from the GCCG upper boundary \(1.557M\) to about \(1.998M\), and introduces a C \(\ell/M=0.90\) sector ending near \(1.948M\), absent from the corresponding GCCG curve. These MCCG throats retain the seed photon sphere and leave its ISCO available for the ISCO-truncated disk. The stable cuts outside the photon sphere found for GCCG instead belong to its different \(\omega=-1.5\) slice. Thus the two EOSs provide different candidates for the optical construction, conditional on the stated material parameters and effective ASG backgrounds. Neither the existence of these sectors nor the retained null barrier determines a black-hole-like image without the separate ray and emission calculation.

The four combinations \((\mathrm B,\mathrm G)\), \((\mathrm B,\mathrm M)\), \((\mathrm C,\mathrm G)\), and \((\mathrm C,\mathrm M)\) are governed by \cref{eq:lapses,eq:static-stresses,eq:gccg-calibration,eq:mccg-calibration,eq:gccg-stability,eq:mccg-stability} on their respective admissible domains. At \(\ell\to0\), both geometries give the same Schwarzschild construction at fixed material scales along any continuous admissible calibration branch. The numerical results evaluate the stated GCCG and MCCG parameter slices on the largest-positive-root branch; they do not exhaust either EOS domain or other calibration branches. The radial analysis concerns spherical shell motion in fixed effective exteriors, without a claim about nonspherical modes, bulk-field perturbations, or a microscopic ASG completion.

\section{Optical appearance of the symmetric configurations}\label{sec:optics}

The stable sectors found above allow two distinct optical constructions: a cut that retains the seed photon sphere and a cut that removes it while retaining the ISCO. We study independent, minimally coupled test radiation in the effective geometries of \cref{eq:metric,eq:lapses}, with a static, transparent shell. The quantum-running prescription therefore enters the optical problem through the lapse and its associated orbital scales. The RG momentum scale \(\varkappa\) is not identified with a photon frequency. No frequency-dependent photon cone, microscopic shell emission or direct coupling of light to the Chaplygin parameters is assumed. This is the same geometric-optics separation used for the transparent magnetic construction of Ref.~\cite{Reboucas2026ExponentialTSW}, now applied to the two quantum-improved backgrounds. The image comparisons below also include static cuts that are unstable in the specified material slices; their radial support is assessed separately.

\subsection{Thin-disk model}\label{subsec:optical-disk}

Each emitting exterior contains a stationary, axisymmetric, geometrically thin disk in the plane \(\theta=\pi/2\). Its self-gravity is neglected. We adopt an optically thin surface-emission prescription: every disk intersection contributes its assigned rest-frame intensity, with no absorption at subsequent intersections or at the throat. This surface prescription does not solve the vertical radiative-transfer problem of a disk of finite thickness. The emitting sides form a set \(\mathcal S\subseteq\{+,-\}\), where \(+\) is the observer's exterior. The unilateral case has \(\mathcal S=\{+\}\), and the bilateral case has \(\mathcal S=\{+,-\}\). In the latter, the two disks have identical radial profiles and the same angular orientation under the identity matching of the spherical coordinates.

The emitters follow the admitted circular timelike geodesics outside the seed ISCO defined by \cref{eq:seed-isco}. Their four-velocity \(u_{\rm em}^{\mu}\), distinguished from the shell tangent, is
\begin{equation}
 u_{\rm em}^{\mu}
 =\frac{(1,0,0,\Omega_{\rm em})}{\sqrt{f-rf'/2}},
 \qquad \Omega_{\rm em}(r)=\sqrt{\frac{f'(r)}{2r}}.
 \label{eq:optical-emitter}
\end{equation}
Here \(\Omega_{\rm em}=d\phi/dt>0\) has dimension \(L^{-1}\); the positive sign fixes the common rotation sense. Normalization requires \(f-rf'/2>0\), and circular motion requires \(f'>0\). We use only the stable outer branch, \(r\geq r_{\mathrm I,X}\), and no emission from plunging material. Even for a face-on observer the gas remains in orbital motion, so transverse Doppler redshift is retained.

We adopt the Gralla--Lupsasca--Marrone (GLM) radial profile~\cite{GrallaLupsascaMarrone2020}, truncated at the ISCO as in Ref.~\cite{Reboucas2026ExponentialTSW}:
\begin{equation}
 I_{\rm em}(r)=
 \begin{cases}
 \displaystyle \frac{I_\star w_{\rm d}}{\sqrt{(r-r_{\mathrm I,X})^2+w_{\rm d}^2}}
 \exp\!\left[-\frac12\operatorname{arsinh}^{2}
 \!\left(\frac{r-r_{\mathrm I,X}}{w_{\rm d}}\right)\right],
 &r_{\mathrm I,X}\leq r\leq r_{\rm d},\\[2pt]
 0,&\text{otherwise}.
 \end{cases}
 \label{eq:optical-emission}
\end{equation}
The positive width \(w_{\rm d}\) and outer disk radius \(r_{\rm d}>r_{\mathrm I,X}\) have dimension \(L\), and \(I_\star>0\) is the emitted specific-intensity scale at the inner edge. We set \(w_{\rm d}=M\) and \(r_{\rm d}=1000M\) for the comparisons below. The profile is taken to be independent of emission direction and locally flat in emitted frequency over the band sampled by the rays. Thus \(I_{\nu,\rm em}(r,\nu)=I_{\rm em}(r)\) within that band, where \(\nu\) denotes frequency; this assumption specifies a fixed-frequency imaging model rather than a broadband accretion spectrum. The inverse hyperbolic sine in \cref{eq:optical-emission} fixes the GLM shape. The finite value at the imposed inner edge and the dark region below it are properties of the selected emission model, rather than discontinuities of the exterior metric.

All optical cuts considered here satisfy \(R_0<r_{\mathrm I,X}\). Consequently the whole prescribed emitting annulus is retained on each illuminated side. A comparison with the corresponding seed black hole uses exactly the same emission law, circular motion, observer and effective lapse. B/C/Schwarzschild comparisons keep \(M,w_{\rm d}/M,r_{\rm d}/M,I_\star\) and the viewing geometry fixed, while allowing each geometry's own ISCO to set its disk edge. This common prescription includes both the shift of the emitting support and the shift of light propagation. Their contributions can be distinguished by the orbital and ray relations below; they should not be attributed to the shell EOS alone.

\subsection{Imaging method}\label{subsec:optical-method}

Let \(\mathsf k^\mu=dx^\mu/d\upsilon\) be a future-directed photon wave vector, with affine parameter \(\upsilon\) normalized to dimension \(L^2\). The conserved Killing energy \(\varepsilon_\gamma=-\mathsf k_t>0\) then has dimension \(L^{-1}\). Spherical symmetry supplies the total angular momentum \(\mathcal J_\gamma\) and its disk-axis component \(\mathcal J_{\gamma z}=\mathsf k_\phi\), both dimensionless in this normalization. Define \(b=\mathcal J_\gamma/\varepsilon_\gamma\geq0\) and \(b_z=\mathcal J_{\gamma z}/\varepsilon_\gamma\), both of dimension \(L\). The null radial equation is
\begin{equation}
 \left(\frac{dr}{d\upsilon}\right)^2
 =\varepsilon_\gamma^2[1-b^2\mathcal U_X(r)],
 \qquad \mathcal U_X(r)=\frac{f_X(r)}{r^2}.
 \label{eq:optical-null}
\end{equation}
The null potential \(\mathcal U_X\), of dimension \(L^{-2}\), is distinct from the shell potential \(\mathcal V\). The stationary point relevant to rays incident from outside is the seed photon sphere already defined by \cref{eq:seed-photon-orbit}. In the orbital plane of a ray, let \(\psi\) be its accumulated angular phase, increasing along backward tracing from \(\psi=0\) at the observer. A monotonic radial leg accumulates
\begin{equation}
 \Delta\psi=\left|\int_{r_a}^{r_b}
 \frac{b\,dr}{r^2\sqrt{1-b^2\mathcal U_X(r)}}\right|.
 \label{eq:optical-phase}
\end{equation}
The radii \(r_a,r_b\) are the endpoints of that leg; turning points solve the vanishing radicand and separate the incoming and outgoing legs. The phase is accumulated continuously through the shell. Continuity of the locally measured photon frequency and of the tangential momentum at a static transparent junction gives identical \(\varepsilon_\gamma,\mathcal J_\gamma,\mathcal J_{\gamma z}\) on the two identical exteriors. The areal radial component changes sign between their outward-pointing charts. There is no additional frequency jump or prescribed specular reflection at the throat.

The static observer lies at \(r_{\rm o}=200M\), \(\theta_{\rm o}=\iota\), \(\phi_{\rm o}=0\), where \(0\leq\iota<\pi/2\) is the inclination to the disk axis and \(f_{\rm o}=f_X(r_{\rm o})\). Its four-velocity is \(u_{\rm o}^{\mu}=f_{\rm o}^{-1/2}(1,0,0,0)\). The exactly edge-on limit requires a finite-thickness transfer model and is excluded. An inward local sky direction has angular separation \(0\leq\alpha<\pi/2\) from the inward radial direction and screen azimuth \(\chi\in[0,2\pi)\). Unit-focal-length stereographic coordinates \((x_{\rm scr},y_{\rm scr})\) are defined by
\begin{equation}
 (x_{\rm scr},y_{\rm scr})
 =2\tan(\alpha/2)(\cos\chi,\sin\chi),
 \qquad b=\frac{r_{\rm o}}{\sqrt{f_{\rm o}}}\sin\alpha.
 \label{eq:optical-screen}
\end{equation}
Both screen coordinates are dimensionless. The finite-observer relation is used without replacing \(b\) by a coordinate distance on the screen. Comparisons can equivalently use the physical angular radius or \(b/M\).

To fix the signed orientation, introduce unit vectors in the auxiliary Cartesian representation of the angular spheres: \(\boldsymbol n_{\rm o}=(\sin\iota,0,\cos\iota)\), \(\boldsymbol e_{\rm h}=(0,1,0)\) and \(\boldsymbol e_{\rm v}=(-\cos\iota,0,\sin\iota)\). Positive screen height is along \(\boldsymbol e_{\rm v}\), toward the projected disk normal. The angular position of the backward ray is
\begin{equation}
 \boldsymbol N(\psi,\chi)=\boldsymbol n_{\rm o}\cos\psi
 +(\boldsymbol e_{\rm h}\cos\chi+\boldsymbol e_{\rm v}\sin\chi)\sin\psi.
 \label{eq:optical-angular-map}
\end{equation}
The unit vector \(\boldsymbol N\) gives the spherical angular position on either copy with identity angular matching. Disk crossings satisfy
\begin{equation}
 \cos\iota\cos\psi+\sin\iota\sin\chi\sin\psi=0,
 \qquad b_z=-b\sin\iota\cos\chi.
 \label{eq:optical-intersections}
\end{equation}
The sign in \(b_z\) refers to the future-directed photon; the integration proceeds backward. At \(\iota=0\), successive intersections have \(\psi=(m-1/2)\pi\), with positive integer crossing order \(m\), and \(b_z=0\). At nonzero inclination their phases depend on \(\chi\), although the radial trajectory at fixed \(b\) remains independent of \(\chi\). Each intersection is assigned its radius \(r_m\) and exterior \(s_m\in\{+,-\}\); intersections at radial-leg boundaries are counted only once.

Frequency transport follows from the measured-frequency ratio \(g_m=\nu_{\rm o}/\nu_{{\rm em},m}\). The contractions with the emitter and observer give
\begin{equation}
 g_m=\frac{-\mathsf k_\mu u_{\rm o}^{\mu}}{-\mathsf k_\mu u_{\rm em}^{\mu}|_{r_m}}
 =\frac{\sqrt{f(r_m)-r_m f'(r_m)/2}}
 {\sqrt{f_{\rm o}}[1-\Omega_{\rm em}(r_m)b_z]}.
 \label{eq:optical-redshift}
\end{equation}
Here \(g_m>0\) is dimensionless, and \(\nu_{\rm o},\nu_{{\rm em},m}\) are the measured observed and emitted frequencies. The denominator is positive on the admitted future-directed circular-emitter rays. The conservation of \(I_\nu/\nu^3\) along a vacuum null ray yields the observed specific intensity
\begin{equation}
 I_{\nu,\rm o}(x_{\rm scr},y_{\rm scr};\nu_{\rm o})
 =\sum_{m:\,s_m\in\mathcal S}
 g_m^3 I_{\nu,\rm em}(r_m,\nu_{\rm o}/g_m).
 \label{eq:optical-intensity}
\end{equation}
Only intersections in the support of \cref{eq:optical-emission} contribute. With the adopted locally flat spectrum, its radial function can be inserted directly into this sum. A bolometric intensity, defined by integration over frequency, would instead carry \(g_m^4\), because the frequency element contributes one further power. We use the specific-intensity observable throughout the present comparisons, consistently with Ref.~\cite{Reboucas2026ExponentialTSW}; bolometric and specific intensities are not interchangeable. For the moving face-on disk, \(g_m^3=[(f-rf'/2)/f_{\rm o}]^{3/2}_{r_m}\). The static-emitter weight \((f/f_{\rm o})^{3/2}\) would describe a different source, even at zero inclination. At nonzero inclination, the \(b_z\) term produces longitudinal Doppler asymmetry while the intersection phases change the illuminated radii.

The celestial-sphere diagnostic is independent of disk illumination. We place a source sphere at \(r_{\rm cel}=1000M\) in each exterior. Bilateral illumination uses \(\mathcal S_{\rm cel}=\{+,-\}\) and the same angular texture on both spheres; unilateral illumination uses \(\mathcal S_{\rm cel}=\{+\}\), with no emission from the opposite sphere. This selection changes the endpoint source, while leaving the geometry and ray trajectories fixed. The observer is equatorial, at \(r_{\rm o}=200M\) and \(\phi_{\rm o}=0\); the edge-on restriction above concerns only thin-disk transfer. We retain the observer-aligned latitude pattern of Ref.~\cite{Reboucas2026ExponentialTSW}. Its positive celestial screen height points along the observer's polar basis vector, opposite to \(\boldsymbol e_{\rm v}\) used for disk display. Thus the celestial endpoint direction is \(\boldsymbol N_{\rm cel}=\boldsymbol N(\psi,-\chi)\) from \cref{eq:optical-angular-map}. The two families of source circles are levels of \(\arcsin(\boldsymbol N_{\rm cel}\cdot\boldsymbol e_{\rm h})\) and \(\arcsin[-\boldsymbol N_{\rm cel}\cdot\boldsymbol e_{\rm v}]\). Their spacing is \(10^\circ\), with black bands of angular half-width \(0.006\) rad. Every panel has a full stereographic field of \(6^\circ\) on each screen axis. Both exteriors use identical cyan, blue, orange and red sectors: in source coordinates along \(\boldsymbol e_{\rm h}\) and \(-\boldsymbol e_{\rm v}\), these occupy quadrants III, IV, I and II, respectively. The colors identify angular directions, rather than intensities or radiation spectra.

Each illuminated sphere additionally carries two identical extended white sources centered at the antipodal directions \(\boldsymbol n_{\rm o}\) and \(-\boldsymbol n_{\rm o}\). Their dimensionless blend weight along a ray is prescribed by
\begin{equation}
 W_{\rm cel}(\psi)=\exp\!\left[-\frac12\left(\frac{\arccos|\cos\psi|}{w_{\rm cel}}\right)^{n_{\rm cel}}\right].
 \label{eq:optical-celestial-source}
\end{equation}
Here \(W_{\rm cel}\in[0,1]\), \(w_{\rm cel}=0.1301668\) rad is the angular width, and \(n_{\rm cel}=3.717993\) is the dimensionless shape exponent. These are prescribed source parameters, independent of the ASG flow and shell EOS. White is blended with the sector color using this weight, after which the black graticule is superposed. The profile has smooth tails rather than a sharp cap boundary; its half-maximum angular radius before the black overlay is approximately \(0.14212\) rad. Its pole images satisfy \(\psi=j_{\rm cel}\pi\), for nonnegative integer \(j_{\rm cel}\): odd orders map to the antipodal pole, and even orders to the observer-aligned pole. The grid remains black at the pole itself, so a narrow dark notch can separate the bright parts of a source image. Neither an individual bright edge nor a black source line identifies a photon orbit.

The prescribed white-source profile is evaluated at each ray endpoint before subpixel averaging. Radial phase maps are interpolated separately on the returning and transmitted branches and checked against complete null trajectories. Celestial images have \(2000\times2000\) pixels; refinement from \(5\times5\) to \(7\times7\) subrays per pixel changes mean channel values by less than \(0.004\). Exact critical directions are treated as limiting rays. The disk and celestial source selections remain independent, and their images are not superposed. Higher-order replicas below this resolution remain unresolved.

For disk display, write \(I_{\rm o}=I_{\nu,\rm o}(\nu_{\rm o})\) at the selected frequency and use
\begin{equation}
 I_{\log}=
 \begin{cases}
 \displaystyle \left[\log_2\!\left(1+\frac{I_{\max}}{I_{\rm o}}\right)\right]^{-1},
 &I_{\rm o}>0,\\[2pt]
 0,&I_{\rm o}=0.
 \end{cases}
 \label{eq:optical-log}
\end{equation}
The common positive reference \(I_{\max}\) is the maximum physical specific intensity over a declared comparison group. This dimensionless, monotonic transformation maps that maximum to one. It is applied after summing the selected exterior contributions. It changes visual contrast but neither the emission profile nor the null trajectories; colors cannot replace comparisons of the untransformed intensities.

The disk images use \(\iota=80^\circ\), \(r_{\rm o}=200M\) and a full \(6^\circ\) stereographic field on each axis. Positive screen height follows \(\boldsymbol e_{\rm v}\) in \cref{eq:optical-angular-map}; it is opposite to the celestial display convention specified above. The two disk matrices share one \(I_{\max}\), taken over all twelve panels, and one cubehelix color scale. For the unilateral/bilateral disk comparison, the two panels share their own reference maximum after summing the selected exterior contributions, with cubehelix representing total intensity in both. Identical disks have the same emission parameters and positive circular rotation in the angular coordinates identified at the shell. Disk images contain \(1600\times1600\) pixels. Radial sampling is concentrated near the critical directions, and pixels whose interpolation cells straddle an emitting-support boundary are evaluated directly from their ray paths. This preserves the sharp ISCO truncation rather than smoothing emission into the excluded region. Refinement of the radial and azimuthal sampling checks the displayed intensity pattern; critical images narrower than the adopted sampling remain unresolved.

\subsection{Imaging results}\label{subsec:optical-results}

The first distinction is determined by the largest null barrier retained outside the cut. For the single exterior maximum present in the adopted parameter ranges, the transmission threshold \(b_{\rm tr,X}\) is
\begin{equation}
 b_{\rm tr,X}^{-2}=\max_{r\geq R_0}\mathcal U_X(r),
 \qquad
 b_{\rm tr,X}=
 \begin{cases}
 b_{\rm ph,X},&R_0\leq r_{\rm ph,X},\\
 R_0/\sqrt{f_X(R_0)},&R_0>r_{\rm ph,X},
 \end{cases}
 \label{eq:optical-threshold}
\end{equation}
where \(b_{\rm ph,X}=r_{\rm ph,X}/\sqrt{f_X(r_{\rm ph,X})}\) is the seed critical impact parameter. All these impact parameters have dimension \(L\). For an observer outside the barrier, \(b<b_{\rm tr,X}\) reaches the opposite exterior, whereas \(b>b_{\rm tr,X}\) returns to the observer's exterior. Critical or exactly tangent directions require a limiting treatment and do not define a finite image pixel by themselves. The corresponding local angular boundary obeys \(\sin\alpha_{\rm tr}=b_{\rm tr,X}\sqrt{f_{\rm o}}/r_{\rm o}\). The second branch in \cref{eq:optical-threshold} is an endpoint barrier at the shell; it is not the photon sphere of an exterior retained after that sphere has been excised.

For \(r_{X,+}<R_0<r_{\rm ph,X}<r_{\mathrm I,X}\), the unilateral disk has an exact seed-black-hole degeneracy under the stated assumptions. Returning rays turn outside the photon sphere and never encounter the cut. The observer-side incoming segment of a transmitted ray coincides with the incoming segment of a captured seed ray. Their differing continuations lie below the disk inner edge on the observer's side, and the other exterior is dark. Thus their complete observer-side sums in \cref{eq:optical-intensity} agree, including emitter motion. In symbols,
\begin{equation}
 I_{\nu,\rm o}^{\{+\},\,\rm shell}(b,\chi;\iota)
 =I_{\nu,\rm o}^{\rm seed}(b,\chi;\iota),
 \qquad R_0<r_{\rm ph,X},\quad 0\leq\iota<\pi/2.
 \label{eq:optical-inner-equality}
\end{equation}
The superscript \(\rm seed\) denotes the black hole with the same \(f_X\), and \(\rm shell\) denotes the matched wormhole. This is an equality of the adopted images, not an equivalence of their global causal structures. It answers the observer-side-disk question for all admitted inclinations. B and C can still differ from Schwarzschild because their seed photon scales, ISCOs and frequency weights differ; retaining the photon sphere does not remove those quantum-background differences.

For \(r_{\rm ph,X}<R_0<r_{\mathrm I,X}\), the impact interval \(b_{\rm ph,X}<b<b_{\rm tr,X}\) changes its trajectory class. A seed ray in this interval turns below \(R_0\) and returns through its emitting exterior, whereas the wormhole ray crosses the shell. The shared incoming segment has identical intensity weights, but selected outgoing observer-side intersections are absent. Since every emitted contribution is nonnegative,
\begin{equation}
 I_{\nu,\rm o}^{\{+\},\,\rm shell}\leq I_{\nu,\rm o}^{\rm seed},
 \qquad b_{\rm ph,X}<b<b_{\rm tr,X}.
 \label{eq:optical-outer-inequality}
\end{equation}
Outside this interval the unilateral images agree, apart from the excluded limiting directions. The inequality is strict only when an omitted outgoing intersection lies in the emitting annulus. It therefore identifies where a difference can occur without claiming that the whole annulus is visibly dimmer at every inclination. The disk-crossing phases in \cref{eq:optical-intersections} determine which screen azimuths actually lose emission. Removing the exterior photon maximum also removes the associated exterior logarithmic winding sequence from the accessible ray family. This statement does not exclude null structures associated with the shell itself.

The stability results supply material-supported examples in both optical regimes. With \(M=d=1\), \(\gamma=1\) and the five deformation values used in the stability figures, \(R_0/M=2.05\) is radially stable for MCCG at \(\omega=-0.5,\beta=2\); it lies below every seed photon sphere and above every outer horizon. The cut \(R_0/M=4.5\) is radially stable for GCCG at \(\omega=-1.5\), above every seed photon sphere and below every ISCO. Each equilibrium uses its own largest-positive-root calibration, held fixed during perturbations. These examples establish stable configurations on either side of the seed photon sphere, without a universal GCCG/MCCG separation.

The B/C image comparisons use the common cuts \(R_0/M=2.42\) and \(3.25\), with \(\ell/M=0,0.4,0.9\), to examine light propagation closer to the photon region. Both cuts remain outside every outer horizon and below every ISCO; the former retains every seed photon sphere and the latter excises it. Their geometric admissibility does not imply stability in the preceding material slices. At the inner cut, MCCG with the stated \(\gamma,\omega,\beta,d\) is stable for \(\ell/M=0,0.4\) in both schemes, but gives \(M^2\mathcal V''_{\mathrm M,0}\simeq-0.00561\) in B and \(-0.20354\) in C at \(0.9\). At the outer cut, the stated GCCG slice is unstable in B for all three columns and in C for the first two; C at \(0.9\) has positive curvature, approximately \(0.02056\). The images therefore include unstable static equilibria as geometric diagnostics. The EOS affects their material viability, while the ray map at fixed \(f_X,R_0\) is independent of that choice.

\begin{table}[t]
\centering
\caption{Seed orbital scales and transmission thresholds for the B/C image comparisons, at equal geometric mass. The last column uses the outer cut \(R_0/M=3.25\); the inner cut \(R_0/M=2.42\) has \(b_{\rm tr,X}=b_{\rm ph,X}\). The zero-deformation row is shared by B and C. Values are evaluated from the full lapses and then rounded.}
\label{tab:optical-scales}
\begin{tabular}{lccccc}
\hline\hline
Geometry & \(\ell/M\) & \(r_{\rm ph,X}/M\) & \(r_{\mathrm I,X}/M\) & \(b_{\rm ph,X}/M\) & \(b_{\rm tr,X}/M\)\\
\hline
Schwarzschild & 0.00 & 3.0000 & 6.0000 & 5.1962 & 5.2405\\
B & 0.40 & 2.9545 & 5.9141 & 5.1490 & 5.2091\\
C & 0.40 & 2.9078 & 5.8262 & 5.1009 & 5.1790\\
B & 0.90 & 2.7405 & 5.5287 & 4.9340 & 5.0896\\
C & 0.90 & 2.4233 & 4.9733 & 4.6234 & 4.9652\\
\hline\hline
\end{tabular}
\end{table}
Table~\ref{tab:optical-scales} shows the inward displacement of the photon sphere and ISCO as the deformation increases, with a larger shift in C at equal \(\ell/M\). For the outer cut, the trajectory-changing interval \(b_{\rm ph,X}<b<b_{\rm tr,X}\) has width approximately \(0.0443M\) in Schwarzschild, \(0.1555M\) in B and \(0.3418M\) in C at \(\ell/M=0.9\). This width specifies the directions in which unilateral disk images can differ from their seed controls; it is not a measured flux deficit. The inner cut is particularly close to the C photon sphere at the largest deformation, lying only approximately \(0.00332M\) below it. It nevertheless remains on the retained-sphere branch of \cref{eq:optical-threshold}.

The origin of this ordering can be isolated perturbatively from \cref{eq:asymptotic}. Let \(a_X\) denote its leading deformation coefficient, with \(a_{\mathrm B}=1\) and \(a_{\mathrm C}=2\). Expanding the existing seed orbit conditions at fixed \(M\) gives
\begin{equation}
 \frac{r_{\rm ph,X}}{M}=3-\frac{5a_X}{18}\left(\frac{\ell}{M}\right)^2
 +O\!\left[(\ell/M)^4\right],
 \quad
 \frac{r_{\mathrm I,X}}{M}=6-\frac{19a_X}{36}\left(\frac{\ell}{M}\right)^2
 +O\!\left[(\ell/M)^4\right].
 \label{eq:optical-orbit-expansion}
\end{equation}
The associated seed critical impact satisfies
\begin{equation}
 \frac{b_{\rm ph,X}}{M}=3\sqrt{3}\left[1-\frac{a_X}{18}
 \left(\frac{\ell}{M}\right)^2+O\!\left[(\ell/M)^4\right]\right].
 \label{eq:optical-impact-expansion}
\end{equation}
These small-deformation relations explain why C produces twice the leading shift of B at equal \(\ell/M\); the table uses the full lapses rather than these truncated series. The shifts inherit the different RG running and the adopted radial scale identification. Their interpretation remains conditional on that effective quantum-background construction. They neither establish a unique ultraviolet prediction nor identify a microscopic light--gravity interaction. The finite-observer angular sizes also contain \(f_{\rm o}\), and intensity weights contain the orbital derivative of \(f\), so a smaller photon scale alone does not establish a universal brightness ordering.

Bilateral disk illumination adds the separate, nonnegative opposite-side sum in \cref{eq:optical-intensity}. It can therefore distinguish the inner-cut wormhole from its seed control even when \cref{eq:optical-inner-equality} holds for unilateral emission. Its visibility still requires transmitted rays to intersect the emitting annulus: a disk does not illuminate every accessible direction, and a central dark region need not imply a horizon. An extended celestial sphere instead tests the endpoint map over all its illuminated angles. The two aligned white sources provide controlled angular references through \cref{eq:optical-celestial-source,eq:optical-angular-map}. Their image positions follow the null map, whereas their widths also depend on the prescribed angular profile and black overlay. These analytical results and calculated scales delimit the image comparisons; no arc contrast, peak intensity or ring width is inferred without evaluating the corresponding source-weighted ray sums.

\begin{figure}[!htbp]
    \centering
    \includegraphics[width=0.98\textwidth]{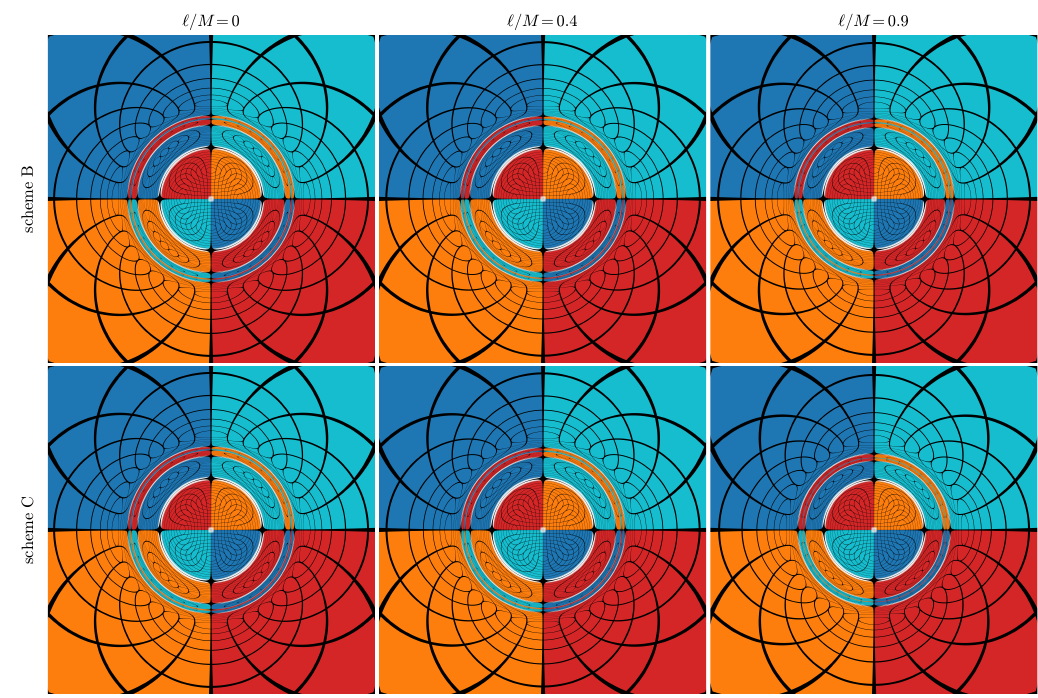}
    \caption{Celestial endpoint maps for a throat at \(R_0/M=2.42\), below the seed photon sphere and above the outer horizon in every panel, with \(M=1\). Rows show scheme B and scheme C; columns have \(\ell/M=0,0.4,0.9\). Both spheres at \(1000M\) carry the same \(10^\circ\) grid, black-band half-width \(0.006\) rad, cyan/blue/orange/red sectors and antipodal white-source profile of Eq.~\eqref{eq:optical-celestial-source}, with black lines superposed. An equatorial static observer at \(200M\) views a common \(6^\circ\) stereographic field per axis. The zero-deformation panels coincide and represent the Schwarzschild-seed wormhole. At this cut, the MCCG slice \(\gamma=1,\omega=-0.5,\beta=2,d=1\) is stable in the first two columns and unstable in the last; the figure compares static geometries rather than exclusively stable shells.}
    \label{fig:asg-celestial-inner}
\end{figure}
Figure~\ref{fig:asg-celestial-inner} has the same basic structure in both rows: a central transmitted image, annular repetitions near the critical boundary, and a returning outer field. The identical first-column panels are an exact control, since both lapses reduce to Schwarzschild at zero deformation. Their colored center and repeated structures already occur in the classical wormhole; neither is by itself a quantum-gravity signature. In particular, the center is illuminated through the second exterior and is not a black-hole shadow. The similarities persist at \(\ell/M=0.4\), where the lapse corrections are relatively small. The B/C distinction becomes clearer in the last column as the compressed annular region contracts more strongly in C.

The angular diameter of the return--transmission boundary decreases from \(2.963^\circ\) at zero deformation to \(2.936^\circ\) in B and \(2.908^\circ\) in C at \(\ell/M=0.4\), and to \(2.813^\circ\) and \(2.636^\circ\), respectively, at \(0.9\). These values follow from \cref{eq:optical-screen,eq:optical-threshold}, rather than from a colored-band width. At fixed cut and source prescription, C therefore produces the larger contraction relative to Schwarzschild. The \(\psi=\pi\) transmitted source image moves much less: its reference diameter changes from \(1.882^\circ\) to \(1.861^\circ\) in B and \(1.842^\circ\) in C at the last column. The unequal shifts change the spacing between the innermost source image and the critical annuli; the panels are not related by a uniform resizing.

The multiple curved black lines are preimages of the two angular grid families. Their folds and repeated sectors follow the accumulated phase in \cref{eq:optical-phase,eq:optical-angular-map}. A retained unstable photon sphere permits logarithmically large phase accumulation as the critical direction is approached, compressing successive grid and white-source images into increasingly narrow annuli. In the C last-column panel, the throat lies very close to that sphere, but does not remove its exterior maximum. Thus the limiting winding hierarchy remains, even if only its broader members are resolved. Its existence is distinct from the visibility of a particular white band, which also depends on the finite source profile and the black notch. No additional count of resolved rings is inferred from increasingly thin pixel structures.

The alternating central colors do not require an inverted palette on the second sphere. The two transverse components of \cref{eq:optical-angular-map} are proportional to \(\sin\psi\); a sign change reverses both and exchanges opposite quadrants. Both sides use the same source texture and signed camera convention, without a subsequent image flip. At a fixed seed, returning rays that turn outside both selected cuts have the same endpoint map in the two figures. The changes inside the boundary instead reflect transmission through different retained radial domains. These bilateral celestial images can reveal the second exterior even though an observer-side disk alone obeys the seed-black-hole equality in \cref{eq:optical-inner-equality} at the inner cut.

\begin{figure}[!htbp]
    \centering
    \includegraphics[width=0.98\textwidth]{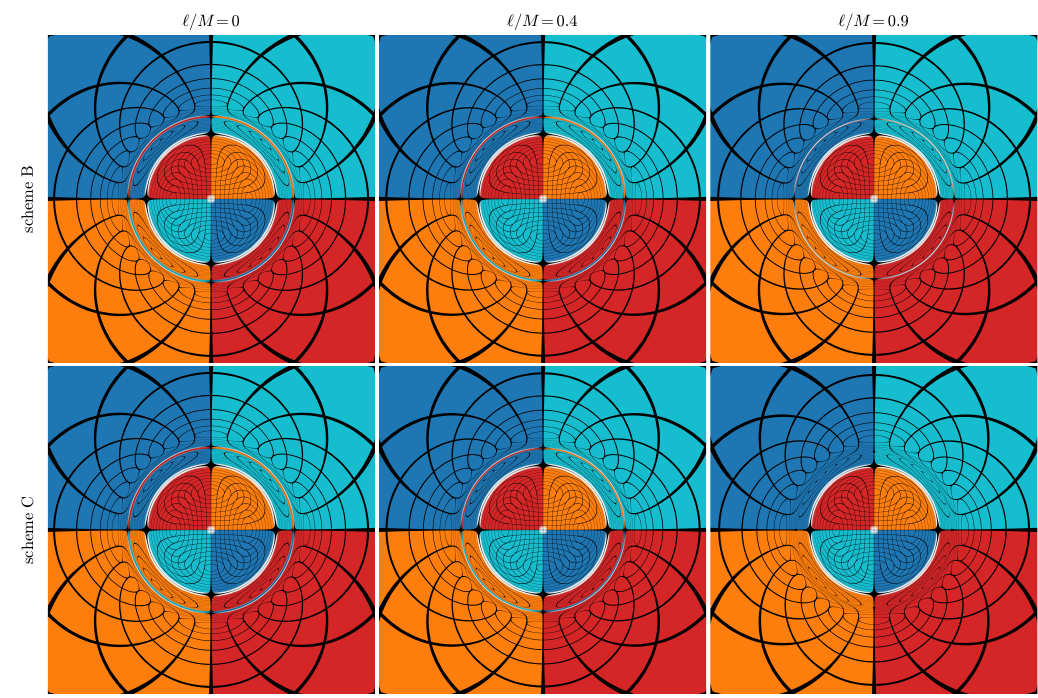}
    \caption{Celestial endpoint maps for \(R_0/M=3.25\), above the seed photon sphere and below the seed ISCO in every panel, with \(M=1\). Rows show scheme B and scheme C; columns have \(\ell/M=0,0.4,0.9\). Observer, source spheres, \(6^\circ\) field, \(10^\circ\) black grid, identical side palettes and antipodal white sources are unchanged from Fig.~\ref{fig:asg-celestial-inner}. The coincident zero-deformation panels are the Schwarzschild-seed wormhole. For GCCG with \(\gamma=1,\omega=-1.5,d=1\), only scheme C at \(\ell/M=0.9\) has positive radial curvature at this cut; the remaining panels are unstable in that material slice. All calibrations use the largest positive root and remain fixed during each shell perturbation.}
    \label{fig:asg-celestial-outer}
\end{figure}
Figure~\ref{fig:asg-celestial-outer} retains the central transmitted field and curved grid images, but removes the exterior photon maximum responsible for the unbounded winding sequence. The transmission boundary now samples the lapse at the throat, through the second branch of \cref{eq:optical-threshold}. Its angular diameter is \(2.988^\circ\) for Schwarzschild, \(2.970^\circ\) and \(2.953^\circ\) for B and C at \(\ell/M=0.4\), and \(2.902^\circ\) and \(2.831^\circ\) at \(0.9\). It is larger than the corresponding inner-cut boundary in every matched panel. The principal transmitted source-image diameter grows to \(2.372^\circ\) in Schwarzschild and remains close to it in B and C at \(0.9\), approximately \(2.359^\circ\) and \(2.346^\circ\). Moving the throat therefore changes the source-image spacing much more than its deformation shifts this principal image.

The most conspicuous B/C distinction concerns a finite phase limit, rather than the loss of all annular structures. In the grazing limit, the total phase from observer to source is approximately \(7.496\) rad for Schwarzschild, \(7.251\) and \(7.044\) rad for B and C at \(\ell/M=0.4\), and \(6.532\) and \(5.998\) rad at \(0.9\). The \(\psi=2\pi\) source order is therefore accessible on both returning and transmitted branches in the first two columns and in the B last-column panel, but not in the C last-column panel. This removes a pair of pole-image centerlines in C while leaving the \(\psi=\pi\) transmitted image. Smooth source tails can still contribute near the missing order; its center is absent, rather than every white pixel there. The \(2\pi\) reference diameters at zero deformation are approximately \(2.973^\circ\) on the transmitted branch and \(3.051^\circ\) on the returning branch. In B at \(0.9\), they approach the boundary at \(2.901^\circ\) and \(2.926^\circ\), making their separation harder to resolve. White features consequently need not disappear at the same parameter value at which two broad colored regions begin to look alike.

Although every outer-cut phase limit is finite, it can exceed \(2\pi\) because the cut remains relatively close to the seed photon region. The surviving curved grid and low-order source images are therefore compatible with excision of the exterior photon sphere. A finite number of broad annuli does not reinstate an exterior logarithmic hierarchy, and the calculation does not exclude every possible null structure associated with the shell itself. Conversely, the denser annuli of Fig.~\ref{fig:asg-celestial-inner} are explained by the retained maximum, not by a denser prescribed source grid: the source and camera are identical in both figures.

The stronger displacement in C traces its greater suppression of the Newton coupling at equal normalized cutoff, propagated through the solution-improved lapse into the null potential and phase. At small deformation, \cref{eq:optical-orbit-expansion,eq:optical-impact-expansion} explain the common near-Schwarzschild appearance and the larger leading C shift. The full lapses determine the finite-deformation phase crossings illustrated here. These are conditional consequences of two coarse-graining prescriptions for quantum gravitational fluctuations combined with the adopted radial scale identification. They do not establish a unique microscopic ASG signature: the angular source pattern is prescribed, the shell material is phenomenological, and no photon dispersion or quantum shell dynamics is included. The stability distinction remains essential for physical interpretation: the outer C last-column image has restoring support in the stated GCCG slice, whereas its inner counterpart is unstable in the stated MCCG slice. This contrast concerns material viability at fixed geometric cuts, rather than an EOS-dependent interaction of the light.

\begin{figure}[!htbp]
    \centering
    \includegraphics[width=0.98\textwidth]{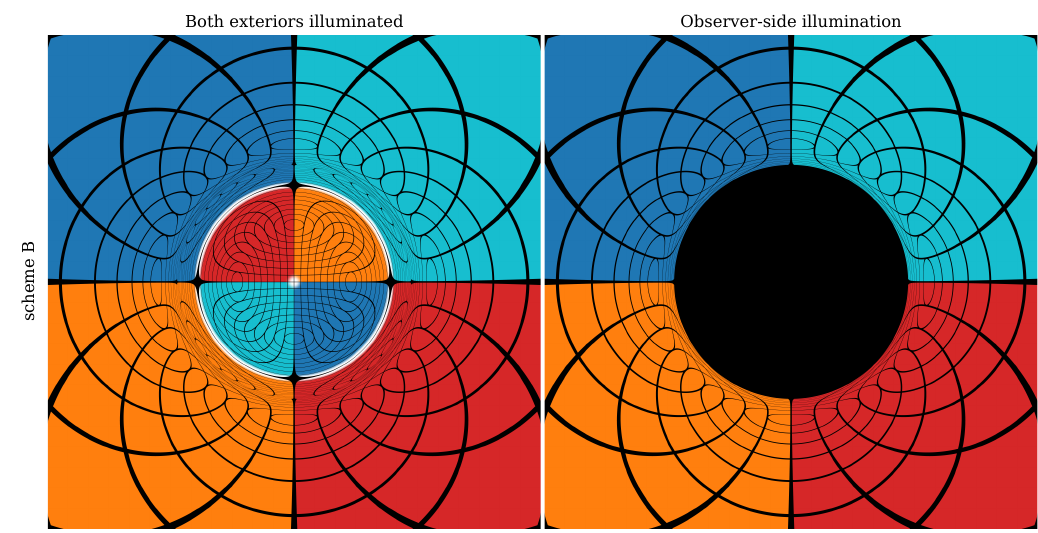}
    \caption{Celestial illumination comparison for scheme B at \(\ell/M=1.20\), \(R_0/M=3.25\), and \(M=1\). Both source spheres are illuminated on the left; only the observer-side sphere emits on the right. The throat lies above the seed photon sphere and below its ISCO. The observer at \(200M\), source spheres at \(1000M\), full \(6^\circ\) stereographic field per axis, \(10^\circ\) grid with black-band half-width \(0.006\) rad, cyan/blue/orange/red palette and white-source profile of Eq.~\eqref{eq:optical-celestial-source} are those of Figs.~\ref{fig:asg-celestial-inner} and~\ref{fig:asg-celestial-outer}. All emission, including the additional white sources, is absent from the opposite sphere in the right panel. The shell has positive radial curvature on the GCCG branch with \(\gamma=1,\omega=-1.5,d=1\), using the largest positive calibration root and fixed material parameters during perturbations.}
    \label{fig:asg-celestial-illumination-b}
\end{figure}
Figure~\ref{fig:asg-celestial-illumination-b} isolates the effect of the emitting exterior in a radially stable configuration. At this deformation, the B seed has \(r_{\mathrm B,+}/M\simeq1.38685\), \(r_{\rm ph,B}/M\simeq2.45055\), and \(r_{\mathrm I,B}/M\simeq5.07123\). Thus the selected throat removes the exterior photon maximum while preserving the whole prescribed ISCO-truncated disk support. The GCCG calibration gives \(M^2\mathcal V''_{\mathrm G,0}\simeq0.01393>0\) for the free material inputs stated in the caption. At the same radius and free inputs, the Schwarzschild control has curvature approximately \(-0.11653\). This change of sign occurs along separately calibrated equilibrium models; it does not hold the material amplitude fixed between different geometries. The selected amplitude is fixed during each local perturbation, as required by \cref{eq:gccg-calibration,eq:linear-motion}.

The two panels have identical null trajectories and the same return--transmission boundary, with \(b_{\rm tr,B}/M\simeq4.98448\) and angular diameter approximately \(2.842^\circ\). Returning rays sample the observer-side sphere in both cases, so their outer grid images coincide. Transmitted rays reach the opposite sphere: its colored sectors and aligned white-source images populate the left panel, whereas their emission is absent on the right. The resulting central black disk is therefore a consequence of source support, rather than an event horizon or a new absorption mechanism at the transparent shell. With the same observer-side celestial sphere, the seed black hole instead captures rays below \(b_{\rm ph,B}/M\simeq4.66569\). Over the interval \(b_{\rm ph,B}<b<b_{\rm tr,B}\), of width approximately \(0.31879M\), its rays return to the illuminated sphere, whereas the wormhole rays reach the dark exterior. A similar central darkness therefore implies neither identical endpoint maps nor identical causal structures. The source choice is applied to each traced direction before pixel averaging, so boundary pixels can mix the two trajectory classes without changing the critical angular scale.

The principal transmitted white-source image has reference diameter approximately \(2.349^\circ\) on the left and disappears when that sphere is darkened. Its position follows the \(\psi=\pi\) endpoint condition, rather than the maximum of the exterior null potential. The grazing phase is finite, approximately \(6.06465\) rad, below \(2\pi\); the \(2\pi\) pole-image centers are therefore absent from both branches. Curved grid lines and a broad white annulus can consequently survive without a retained exterior photon sphere, while their removal by changing illumination is distinct from the removal of an allowed ray order by changing geometry. Faint smooth source tails, if visible away from a pole center, remain governed by \cref{eq:optical-celestial-source}.

Relative to the Schwarzschild-seed outer-cut control, whose transmission-boundary diameter is approximately \(2.988^\circ\), the B quantum-improved background contracts this scale while still admitting the selected stable shell. Relative to B at \(\ell/M=0.9\), the larger deformation also lowers the grazing phase from approximately \(6.532\) rad to below \(2\pi\). These changes follow the running-coupling lapse and retained null potential; darkening the opposite source at fixed deformation does neither. The value \(\ell/M=1.20\) lies in B's additional horizon-admitting domain and has no C black-hole-seed counterpart at equal deformation within the adopted construction. The present comparison therefore tests illumination in one prescribed quantum-improved geometry, without confusing the extended B range with an equal-parameter B/C difference.

In this stable configuration, the ASG flow and radial scale identification determine the effective background, the phenomenological GCCG response supplies radial restoring support, and the external sources determine which allowed rays are illuminated. Neither the black region nor the white annulus alone identifies microscopic quantum gravity. The two source geometries probe different emission supports: a celestial sphere illuminates angular endpoints, whereas an observer-side disk contributes only at its admitted intersections and obeys the conditions of \cref{eq:optical-inner-equality,eq:optical-outer-inequality}.

\begin{figure}[!htbp]
    \centering
    \includegraphics[width=0.98\textwidth]{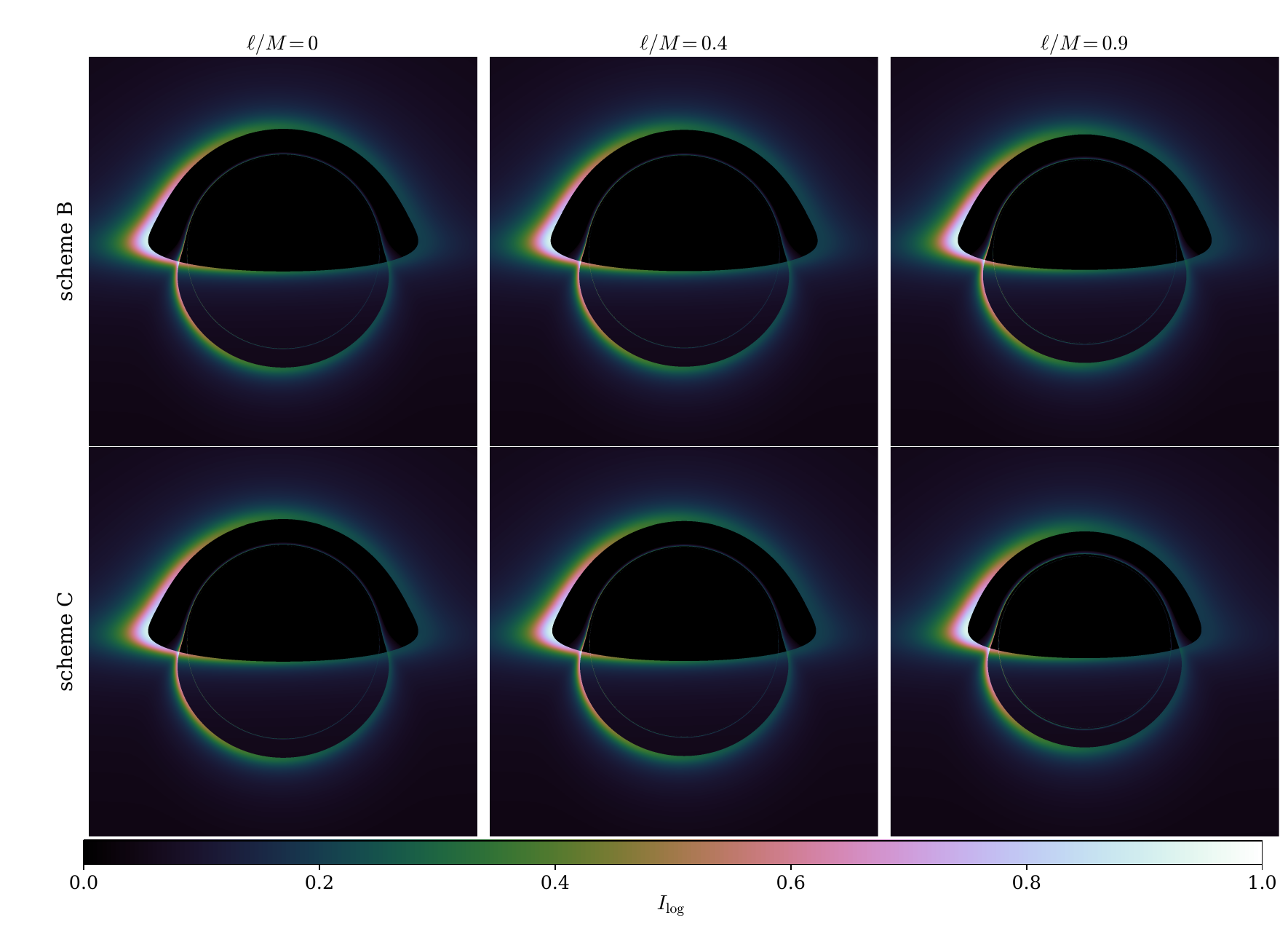}
    \caption{Inclined observer-side disk images for \(R_0/M=2.42\), below every seed photon sphere, with \(M=1\). Rows show scheme B and scheme C; columns have \(\ell/M=0,0.4,0.9\). The disk follows \cref{eq:optical-emission}, from each seed's own ISCO to \(r_{\rm d}=1000M\), with \(w_{\rm d}=M\); the opposite exterior is dark. The static observer has \(r_{\rm o}=200M\) and \(\iota=80^\circ\), with a full \(6^\circ\) field per axis. Positive height follows the disk frame of \cref{eq:optical-angular-map}. Colors show the specific intensity through \cref{eq:optical-log}, using one reference maximum over the twelve disk panels. The identical first-column images are the Schwarzschild-seed control. The material classification is that of Fig.~\ref{fig:asg-celestial-inner}: the first two columns are stable in the stated MCCG slice, whereas the last is unstable.}
    \label{fig:asg-disk-inner}
\end{figure}
Figure~\ref{fig:asg-disk-inner} shows a common bright crescent on the left, a bent image of the far side of the disk above the dark central region, and fainter internal lensed features. These structures are already present in the Schwarzschild control. The strong horizontal asymmetry follows emitter motion: on the left, the signed angular momentum \(b_z\) is positive, reducing the denominator in \cref{eq:optical-redshift} and increasing the observed frequency and specific-intensity weight. The sharp emitting edge is imposed at the ISCO; it is not a discontinuity in the lapse or evidence of a singular shell crossing. The logarithmic display makes weak secondary contributions visible alongside the bright crescent, without assigning equal physical intensity to similarly broad colored structures.

The first two columns remain close in appearance, while the largest deformation reveals a greater contraction in C than in B. Table~\ref{tab:optical-scales} gives an ISCO reduction relative to Schwarzschild of approximately \(7.9\%\) in B and \(17.1\%\) in C at \(\ell/M=0.9\), compared with reductions of \(5.0\%\) and \(11.0\%\) in the seed critical impact parameter. The emitting edge and null critical scale therefore move by different fractions. This accounts for changes in their relative spacing, rather than a rigid resizing of the whole panel. The same GLM prescription is used in every geometry, but its inner edge follows that geometry's ISCO. In addition, \(f_X\) and its derivative change the orbital frequency and frequency transport. These simultaneous changes prevent a universal brightness ordering from being inferred from the smaller C photon scale alone.

Under the unilateral source assumptions, every panel also represents the corresponding seed-black-hole image through \cref{eq:optical-inner-equality}, despite the different continuation of transmitted rays. This degeneracy includes the moving disk and the adopted inclination. In particular, the small separation between the C photon sphere and the throat at the largest deformation does not break it while the sphere remains outside the cut. A dark central region in this disk diagnostic cannot establish the presence of a horizon. Conversely, the colored transmitted center in Fig.~\ref{fig:asg-celestial-inner} is supplied by its opposite-side celestial source, absent from the present disk calculation. Their different structures follow from the emission supports and screen orientations specified above.

\begin{figure}[!htbp]
    \centering
    \includegraphics[width=0.98\textwidth]{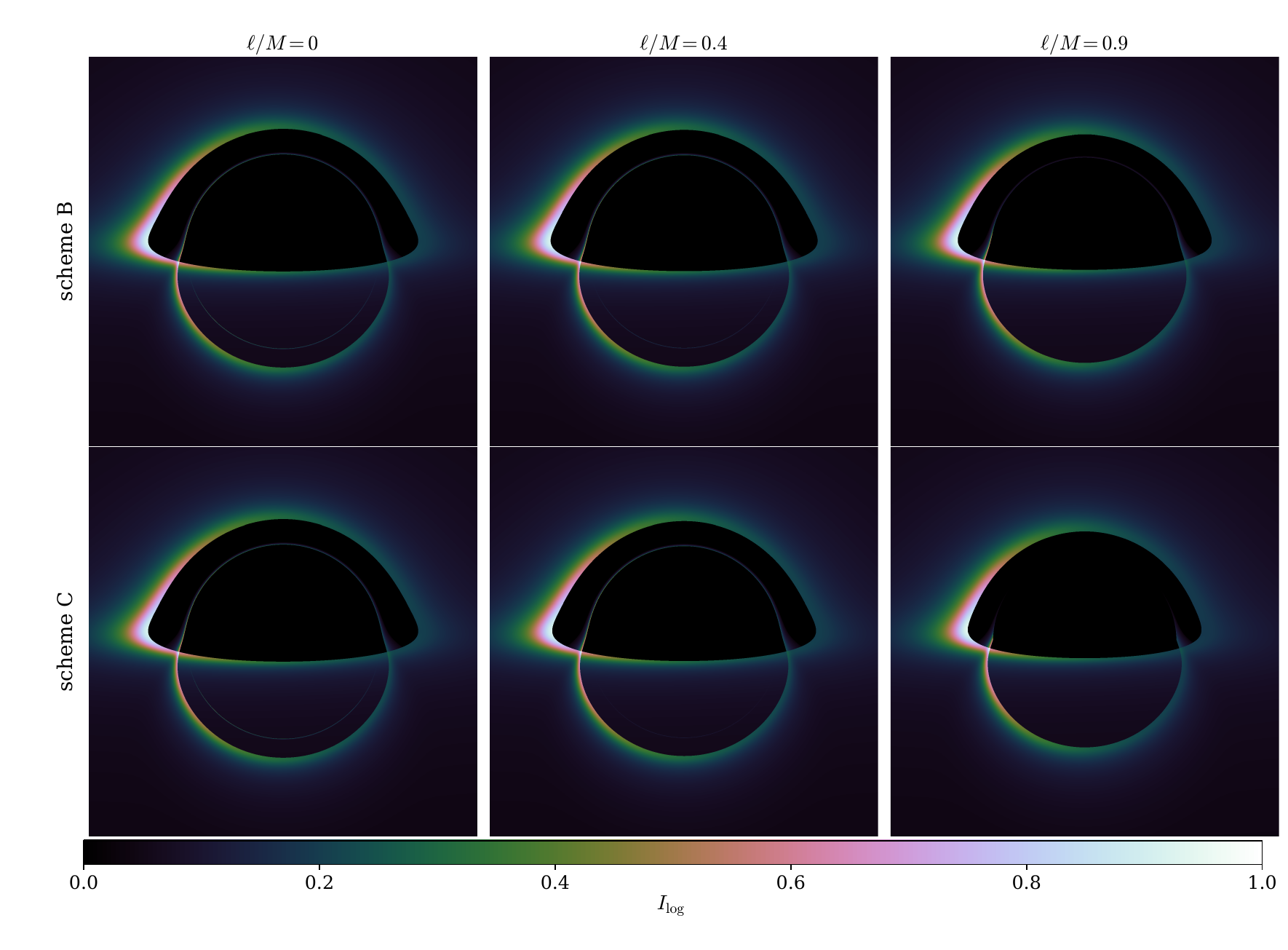}
    \caption{Inclined observer-side disk images for \(R_0/M=3.25\), above every seed photon sphere and below every seed ISCO, with \(M=1\). Rows show scheme B and scheme C; columns have \(\ell/M=0,0.4,0.9\). Disk support, observer position, \(80^\circ\) inclination, signed disk frame, \(6^\circ\) field, cubehelix scale and the twelve-panel normalization are those of Fig.~\ref{fig:asg-disk-inner}. Thus matching panels in the two disk figures differ only in throat position. The material classification is that of Fig.~\ref{fig:asg-celestial-outer}: only C at \(\ell/M=0.9\) is stable in the stated GCCG slice.}
    \label{fig:asg-disk-outer}
\end{figure}
Figure~\ref{fig:asg-disk-outer} preserves the broad Doppler crescent and much of the primary disk image, while weakening or removing internal lensed contributions. The change is clearest in the C last-column panel. For each fixed seed and deformation, both selected cuts lie inside the same disk inner edge, so moving the throat does not remove any part of the prescribed emitting annulus. It instead changes the paths that can return through that annulus. Over \(b_{\rm ph,X}<b<b_{\rm tr,X}\), rays that would turn in the seed geometry cross the wormhole throat and enter its dark exterior. Their shared incoming emission remains, but outgoing observer-side contributions can be lost, as established by \cref{eq:optical-outer-inequality}. The resulting change need not be a completely black circular mask: a transmitted ray can still encounter the observer-side disk on its incoming segment.

The wider trajectory-changing interval in C, quantified in Table~\ref{tab:optical-scales}, allows a larger departure from its seed control than the narrower interval in B at the same deformation. This geometric opportunity is realized only at azimuths with an omitted emitting intersection; it is not an azimuth-independent flux deficit. Direct ray sums inside the interval confirm losses at selected azimuths, while directions outside it retain the seed contributions. Since both disk figures use a common physical intensity reference, their differences are not produced by separate panel normalizations. Excision also removes the retained exterior maximum that supports logarithmically increasing winding. The remaining low-order disk features are compatible with this finite exterior phase range; their number at the displayed resolution is not a count of all possible null structures at the shell.

The B/C ordering connects these optical changes to the quantum origin of the backgrounds. Their different coarse-graining prescriptions for gravitational fluctuations produce different running couplings and, after the adopted radial scale identification, different \(f_X\). At equal \(\ell/M\), C shifts the orbital and critical scales more strongly and increases the interval in which excision changes a returning seed ray into a transmitted one. Emission and propagation are then calculated on these effective backgrounds with classical geometric optics. The panels therefore compare conditional optical consequences of quantum-improved geometries; they do not describe a direct quantum interaction of photons with the Chaplygin gas or establish a unique observational signature of ASG.

The stability assignments also distinguish an allowed static image from a material-supported equilibrium. In the chosen GCCG slice, the C last-column outer cut has positive radial curvature, whereas its inner-cut counterpart is unstable in the chosen MCCG slice. The optical change in that pair can therefore accompany a restoring shell response on the outer branch. The other outer-cut panels remain useful geometric controls but are not stable in the specified material slice. Together with the stable inner and outer examples already identified, this comparison connects stability to the available throat positions without making a universal GCCG/MCCG claim. At fixed \(f_X\) and \(R_0\), changing the phenomenological EOS cannot itself alter the ray sums.

\begin{figure}[!htbp]
    \centering
    \includegraphics[width=0.98\textwidth]{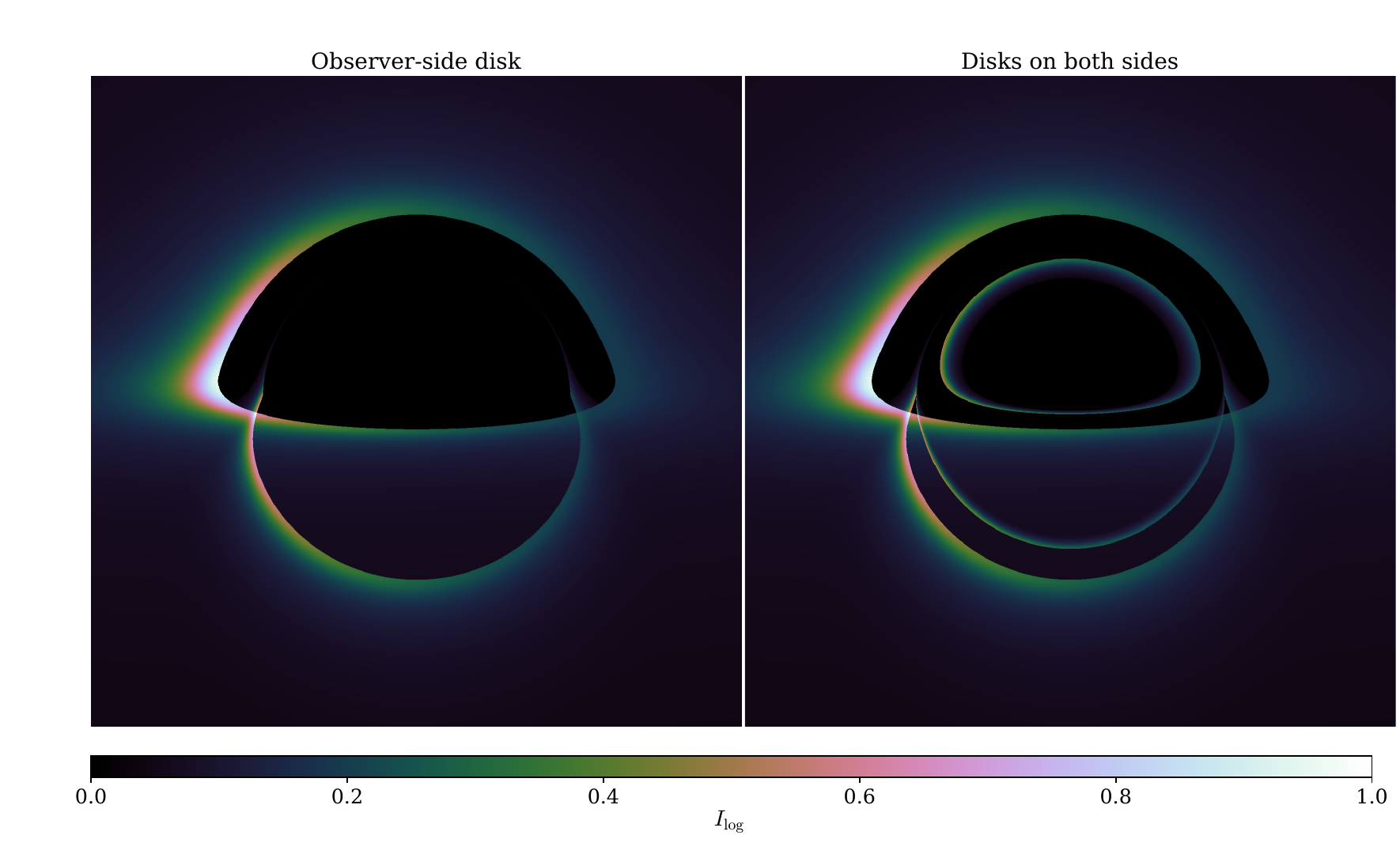}
    \caption{Observer-side disk alone (left) and identical disks in both exteriors (right), for scheme B at \(\ell/M=1.20\), \(R_0/M=3.25\) and \(M=1\), using the geometry of Fig.~\ref{fig:asg-celestial-illumination-b}. The throat lies above the seed photon sphere and below its ISCO. The observer is at \(200M\), with \(\iota=80^\circ\) and a full \(6^\circ\) stereographic field per axis. Each illuminated disk has the GLM profile of \cref{eq:optical-emission}, width \(w_{\rm d}=M\), inner edge at its own ISCO and outer edge at \(1000M\), with the disk frame and rotation used in Figs.~\ref{fig:asg-disk-inner} and~\ref{fig:asg-disk-outer}. Both panels use one pairwise maximum in \cref{eq:optical-log} and the same cubehelix palette for total specific intensity; hue does not identify an emitting exterior. The shell is radially stable in the GCCG slice \(\gamma=1,\omega=-1.5,d=1\), with the largest positive calibration root fixed during perturbations.}
    \label{fig:asg-disk-illumination-b}
\end{figure}
Figure~\ref{fig:asg-disk-illumination-b} changes the illuminated exteriors while retaining the geometry and every observer-side emitting contribution. The broad approaching-side crescent and the returning outer field are the same in both panels. Activating the second disk adds internal arcs, including emission inside part of the dark region of the unilateral image. These features follow intersections with matter in the opposite exterior. The shared cubehelix scale maps the summed specific intensity, so the new arcs are identified with that source through the separated contributions in \cref{eq:optical-intensity}, rather than through a distinct source color. Their addition is nonnegative before the logarithmic transformation; it cannot dim an existing observer-side contribution in the adopted transparent, nonabsorbing model.

Only rays that cross the shell can acquire this extra emission. Returning rays never visit the second disk, which explains the unchanged exterior pattern. Transmission alone is insufficient, however: the continued ray must intersect the emitting annulus beyond the opposite ISCO. Thus a disk illuminates a smaller set of transmitted directions than the extended celestial sphere in Fig.~\ref{fig:asg-celestial-illumination-b}. The bilateral disk image consequently retains a dark central region. At the exact radial direction, the ray has no equatorial crossing at the adopted inclination, and the finite annular support excludes further directions. This darkness is compatible with passage into the second exterior and does not imply capture by a horizon. The source geometry accounts for the difference between the internal arcs here and the celestial grid and aligned-source images.

The additional curved structures do not require restoration of the excised seed photon sphere. They are permitted disk images within the finite phase range of this outer-cut geometry. The grazing-phase restriction already established for Fig.~\ref{fig:asg-celestial-illumination-b} still applies, while the inclined disk selects particular crossing phases through \cref{eq:optical-intersections}. Its emitting edge and Doppler factors determine the position and contrast of the added arcs. These finite-order disk images do not identify additional photon spheres or restore the excised exterior logarithmic sequence. Bilateral illumination also removes the premise of the unilateral seed-image comparison: the separate opposite-side term is present in the wormhole image and has no counterpart in a black hole supplied only with the observer-side disk. Within the trajectory-changing interval of \cref{eq:optical-outer-inequality}, the added opposite-side emission competes with the lost seed returning contributions, so the unilateral inequality does not impose a universal brightness ordering on the bilateral image.

This source comparison takes place on the same quantum-improved B background that admits the restoring GCCG equilibrium discussed with Fig.~\ref{fig:asg-celestial-illumination-b}. The ASG flow and radial scale identification set its null map, orbital edge and frequency weights; changing illumination at fixed background changes none of those quantities or the radial curvature. The established stability contrast with the Schwarzschild control concerns separately calibrated equilibria at the same free material inputs. The present pair isolates the radiation from a second exterior within that supported geometry, rather than a new quantum interaction of light with the shell. The B/C comparison remains the one at shared admissible deformations in the preceding matrices; B's extended horizon-admitting range is not a matched C control. Additional inner emission can test source access through the throat, but its appearance alone is not a unique signature of ASG.

\section{Conclusion}\label{sec:conclusion}
The two ASG prescriptions produce different optical scales and radial restoring sectors within the same symmetric thin-shell construction. Their physical distinction originates in the coarse graining of quantum gravitational fluctuations, carried into the exterior through the running Newton coupling and the adopted IR scale identification. At equal mass and normalized deformation, C suppresses the coupling more strongly than B and moves the outer horizon, seed photon sphere and ISCO farther inward. Its leading small-deformation orbital and critical-impact shifts are twice those of B. These differences are consequences of the specified quantum-improved backgrounds; their interpretation retains the effective scale prescription and the constant-\(G_0\) junction description. Both prescriptions recover the same Schwarzschild limit.

The disk comparison establishes that throat topology and observed brightness need not have a one-to-one correspondence. When the cut lies below the seed photon sphere and emission begins at the ISCO only in the observer's exterior, the complete image equals that of the black hole with the same lapse. This equality includes orbital motion and every admitted inclination. B and C can nevertheless differ from Schwarzschild through their own emitting edges, light deflection and frequency transport. A cut between the photon sphere and ISCO changes the comparison: some rays that return through the seed disk instead cross the throat, losing outgoing observer-side emission. The effect appears only where an omitted intersection lies within the emitting annulus. It preferentially changes internal lensed contributions while much of the primary Doppler crescent survives.

Celestial illumination makes the retained null structure and the prescription dependence more explicit. Inner cuts preserve the exterior photon maximum and the associated increasing winding near the critical direction. Outer cuts have a finite grazing phase, yet still support curved grid images and low-order annuli. The white-source images trace the angular endpoint map, so their existence is controlled by the accessible phase as well as their prescribed angular profile. For \(R_0/M=3.25\) and \(\ell/M=0.9\), B still reaches the \(2\pi\) pole-image order, whereas C does not. This is a qualitative difference within an otherwise similar appearance, accompanying the stronger contraction of the critical scales in C. Removing the exterior photon maximum therefore changes both the winding hierarchy and the available source-image orders, without requiring all visible annuli to disappear.

Illuminating the opposite exterior provides a second, independent way to reveal transmission. Its celestial sphere fills directions that are dark with an observer-side source alone; an opposite-side disk contributes only where a transmitted ray intersects its emitting annulus. The latter adds internal arcs while preserving a residual central dark region. In the GCCG-supported B example at \(\ell/M=1.20\) and \(R_0/M=3.25\), these extra arcs occur on the same radially stable background as the unilateral image. Their addition changes neither the trajectories nor the shell response. The example belongs to B's additional horizon-admitting range, while matched B/C distinctions are established at shared deformations. A dark center and a bright annulus consequently require interpretation through their emitting support and ray paths, rather than identification with an event horizon or an additional photon sphere.

Radial stability supplies the material selection behind these optical regimes. On the largest-positive-root branch and for the sampled exponents, GCCG at \(\omega=-1.5\) admits stable throats beyond the seed photon sphere, including the interval below the ISCO. The stable MCCG sectors found at \(\omega=-0.5\), with \(\beta=1,2,3\), instead lie below their respective seed photon spheres. This contrast is conditional on the sampled material parameters: GCCG also has inner sectors, and the MCCG boundary \(\beta=0\) recovers GCCG. At matched free inputs, C shifts restoring boundaries farther inward, but can remove stability at a common radius while opening it closer to the horizon. The comparisons with Schwarzschild likewise show both gained and lost restoring configurations, depending on radius and material inputs. The classification is local and radial, and the optical matrices also contain static cuts that are unstable in the stated material slices.

The Chaplygin laws enter this result as phenomenological surface constitutive relations. A fixed density normalization and ordinary real rational powers preserve their stated form at negative surface density; the selected equilibrium amplitude is then held fixed during each perturbation. Comparing different backgrounds recalibrates that amplitude, so their stability curves describe families of materials rather than the motion of one fluid with varying coefficients. Negative surface density persists, including where the tangential NEC holds. Quantum running changes the geometry and the stresses required by the junction, while the gas determines the local pressure response; neither role supplies a microscopic quantum realization of the supporting layer. At fixed geometry and throat radius, the stationary optical map remains independent of the gas parameters. The combined analysis relates the optical consequences of gravitational running to the equilibria admitted by a specified surface response.

Following the radially supported branches through changes in viewing inclination and disk support would extend the present comparison to source conditions that can compete with the B/C optical differences. The phase thresholds and lost-return mechanism identified here provide specific quantities against which those changes can be assessed. A physically calibrated relation between the RG scale assignment and the mass of the object would further connect the dimensionless deformations to quantum-gravity scales. Together, these developments would sharpen the distinction between the optical consequences of gravitational running, the radial support furnished by the surface material and the illumination supplied by the two exteriors.

\begin{acknowledgments}
Edson Otoniel would like to thank Fundação Cearense de Apoio ao Desenvolvimento Científico e Tecnológico through grant BP6-0241-00335.01.00/25.
\end{acknowledgments}

\bibliographystyle{apsrev4-2}
\bibliography{references}

\end{document}